\documentclass[draft]{agujournal2019}
\usepackage{url} 
\usepackage[inline]{trackchanges} 
\usepackage{soul}

\draftfalse

\usepackage{amsmath}

\newcommand{\mbf}{\boldsymbol}
\newcommand{\bnabla}{\mbf{\nabla}}

\newcommand{\p}{\partial}

\newcommand{\ddroit}[2]{\ensuremath{\dfrac{\mathrm{d}{#1}}{\mathrm{d}{#2}}}}
\newcommand{\bu}{\mbf{u}}

\newcommand{\ie}{\textit{i.e.}~}
\newcommand{\mean}[1]{\left\langle #1 \right\rangle}

\journalname{Journal of Geophysical Research: Planets}

\begin{document}

\xdefinecolor{brickred}{rgb}{0.8, 0.25, 0.33}
\xdefinecolor{RED}{named}{black}
\xdefinecolor{RED2}{named}{black}

%
%


\title{Internally heated porous convection: an idealised model for Enceladus' hydrothermal activity}

\authors{Thomas Le Reun \affil{1} and Duncan R. Hewitt \affil{2} }

\affiliation{1}{DAMTP, University of Cambridge, Wilberforce Road, Cambridge CB3 0WA, UK}
\affiliation{2}{Department of Mathematics, University College London, UK }

\correspondingauthor{Thomas Le Reun}{tl402@cam.ac.uk}

\begin{keypoints}
\item We carry out numerical and theoretical analysis of an idealised model of tidally driven hydrothermal activity inside Enceladus.
\item With numerical and theoretical analysis, we explore the flow that develops in a porous core with volumetric heating.
\item Our model allows us to predict typical temperature, velocity and heat flux anomalies at the bottom of Enceladus' subsurface ocean.
\end{keypoints}

\begin{abstract}
Recent \textcolor{RED}{planetary} data and geophysical modelling suggest that hydrothermal activity is ongoing under the ice crust of Enceladus, one of Saturn's moons. 
According to these models, hydrothermal flow in the porous, rocky core of the satellite is driven by tidal deformation that induces dissipation and volumetric internal heating.
Despite the effort in the modelling of Enceladus' interior, systematic understanding---and even basic scaling laws---of internally-heated porous convection and hydrothermal activity are still lacking.
In this article, using an idealised model of an internally-heated porous medium, we explore numerically and theoretically the flows that develop close and far from the onset of convection.
In particular, we quantify heat-transport efficiency by convective flows as well as the typical extent and intensity of heat-flux anomalies created at the top of the porous layer.
With our idealised model, we derive simple and general laws governing the temperature and hydrothermal velocity that can be driven in the oceans of icy moons.  
In the future, these laws could help better constraining models of the interior of Enceladus and other icy satellites. 
\end{abstract}

\section*{Plain Language Summary}

\textcolor{RED}{
Enceladus, one of Saturn's icy moons, is known from planetary data to be the site of ongoing hydrothermal activity. 
According to recent modelling, this activity is driven by tidal distortion throughout its porous rocky core, which causes friction and induces volumetric heating. 
As subsurface water penetrates through the core, it warms, rises, and returns into the ocean through localised hotspots. 
We introduce an idealized model of this hydrothermal circulation in order to understand the formation of hot spots, their typical size and their activity. 
We find that the hydrothermal flow in the porous core of Enceladus is about a few centimetres per year and is thus much slower than circulations in the Earth's ocean crust. %
As a result, the timescale for hotspot activity variations is as long as a few million years. Despite the slowness of the circulation, we predict that it drives oceanic plumes with velocity of the order of one centimetre per second.    
}


\section{Introduction} \label{sec:intro}

Enceladus, a 500 km-diameter icy satellite orbiting Saturn, has drawn a lot of attention since the first flybys operated by the Cassini probe in 2005.
Pictures and \textit{in situ} astrochemical measurement have revealed the presence of a water-vapour and ice plume ejected into outer space.
It emerges along fractures in the ice crust at the south pole of Enceladus and is associated with a large heat-flux anomaly of 12.5 GW \textcolor{RED}{\cite{spencer_cassini_2006,spencer_plume_2018}}.
Subsequent analyses have revealed that the ejected material contains silicate particles of nanometric size whose chemistry indicates that the water contained in the plume has been previously hot, liquid, and in contact with silicate rocks \cite{hsu_ongoing_2015,sekine_high-temperature_2015}.
Enceladus' plumes have since then been interpreted as evidence for hydrothermal activity occurring below the ice crust of Enceladus.
This is a surprising implication because, unlike the Earth, Enceladus has radiated away all its initial heat, and its small size makes internal heating by radiogenic elements insufficient to explain the abnormal heat flux \cite{nimmo_ocean_2016,choblet_powering_2017}.
Building on the recent study of \citeA{lainey_new_2017},   
\citeA{choblet_powering_2017} have recently proposed a self-consistent model to explain the hydrothermal activity based on internal heating by tides in Enceladus' water-saturated porous core. 
This model relies on recent findings regarding the interior of Enceladus. 
Underneath its ice crust, this satellite comprise a global subsurface ocean, with thickness varying from  30 to 50 km \cite{thomas_enceladuss_2016-1}. 
Below lies a core made of rocky material that remains undifferentiated and uncompacted owing to to the weakness of Enceladus' gravity field 
\cite{roberts_fluffy_2015,choblet_powering_2017}.
The core is thus permeated with the water of the ocean; \citeA{choblet_powering_2017} estimate that the porosity ranges from 20 to 30\% for a water-filled rocky core. 
Lastly, intense internal heating \cite{lainey_new_2017} due to tidally-induced deformation and friction heats the water and creates a porous flow with hot and narrow upwelling zones, possibly leading to hot spots of water flowing into the ocean \cite{choblet_powering_2017}.
\textcolor{RED}{Hydrothermal convection with internal heating is only a relatively recent feature of thermal evolution models  \cite{travis_whole-moon_2012,travis_keeping_2015}, and it is in general driven by radiogenic heating or serpentinisation rather than tidal deformation \cite{nimmo_ocean_2016}.}
\textcolor{RED}{
However, interior models of icy moons deal with very poorly constrained parameters, for instance the permeability of the core for which a range of orders of magnitudes is plausible \cite{travis_keeping_2015,choblet_powering_2017}. 
Despite systematic studies covering a wide range of parameters \cite{choblet_powering_2017}, general scaling laws predicting the size and intensity of heat-flux anomalies, the typical temperature or hydrothermal velocity and their dependence to physical parameters are still lacking. 
By investigating a basic model for internally-heated porous convection, we aim to derive these scaling laws, which could prove useful to better constrain the planetary data available for Enceladus or to build thermal-evolution models of icy moons more generally \cite{travis_whole-moon_2012,travis_keeping_2015}. 
}

%

%
In the present article, we \textcolor{RED}{thus} explore systematically internally-heated porous convection close to and far from the onset of motion with numerical simulations and mathematical analysis.
We use an idealised two-dimensional Cartesian model of a water-saturated porous layer with internal heating in order to reduce the complexity of the system as much as possible while retaining the key physical ingredients, \textcolor{RED}{which are internal heating and an open-top boundary}. 
%
%
This kind of approach has a long history of use in convection studies.
The canonical model to study heat transport by convection is the Rayleigh-B\'enard set-up (a confined porous layer heated from below and cooled from the top) which has received a significant amount of study  \cite{otero_high-rayleigh-number_2004,hewitt_ultimate_2012,hewitt_high_2014,
hewitt_stability_2017-2}.
The more closely related case of Earth-like hydrothermal systems with a bottom heat flux and open top boundary has also been widely studied (see for instance \citeA{fontaine_two-dimensional_2007,coumou_structure_2008,coumou_phase_2009}).
However, the results of these investigations are unlikely to apply to tidally-driven hydrothermal circulation because of \textcolor{RED}{either unsuitable} boundary condition \textcolor{RED}{or} the nature of the heat source.
Very few systematic experimental and numerical studies have been devoted to internally-heated porous convection.
Those that have are focused mostly on the onset of motion and average heat transport 
\cite{buretta_convective_1976,
nield_onset_2013,hardee_natural_1977,
kulacki_hydrodynamic_1975}.
Hence, these studies do not allow the derivation of scalings governing, for instance, the typical extent of upwelling zones or the associated thermal anomalies and fluid velocities, in the case of tidally-driven hydrothermal activity.
That is our aim here.%

\textcolor{RED}{In common with numerous convection set-ups, we find that the intensity of heat-transporting motion is characterised by only one dimensionless number, the Rayleigh number, noted $Ra$, which increases with volumetric heat production, permeability and core radius.
Performing numerical simulations and asymptotic analysis, we find that the typical size of thermal anomalies is proportional to $Ra^{-1/2}$, owing to a balance between advection, heat production and advection.
As a consequence, the plumes driven in the ocean by thermal anomalies have a buoyancy scaling like $Ra^{3/2}$.
When quantified for ranges of parameters that are expected for Enceladus, we predict typical Darcy fluxes in the core of at most $10$ cm per year, while hydrothermal velocities are expected to reach about $1$ cm.s$^{-1}$. 
%
}
This paper is organised as follows. 
A first part is devoted to introducing our idealised model for an internally-heated, saturated porous layer and identifying the relevant dimensionless parameters.
We then carry out a stability analysis to determine the conditions under which convection happens.
Afterwards, we describe and analyse numerical simulations of internally-heated porous convection, focusing in particular on the structure of the flow and the associated thermal anomalies.
Lastly, we apply the laws derived from our idealised model to Enceladus to quantify the temperature anomalies and the typical hydrothermal velocities that can be induced in its ocean.

\section{A simple model for the interior of icy moons}

\subsection{The model and its governing equations}
\label{sec:model_introduction}

\begin{figure}
\centering
\includegraphics[width=0.4\linewidth]{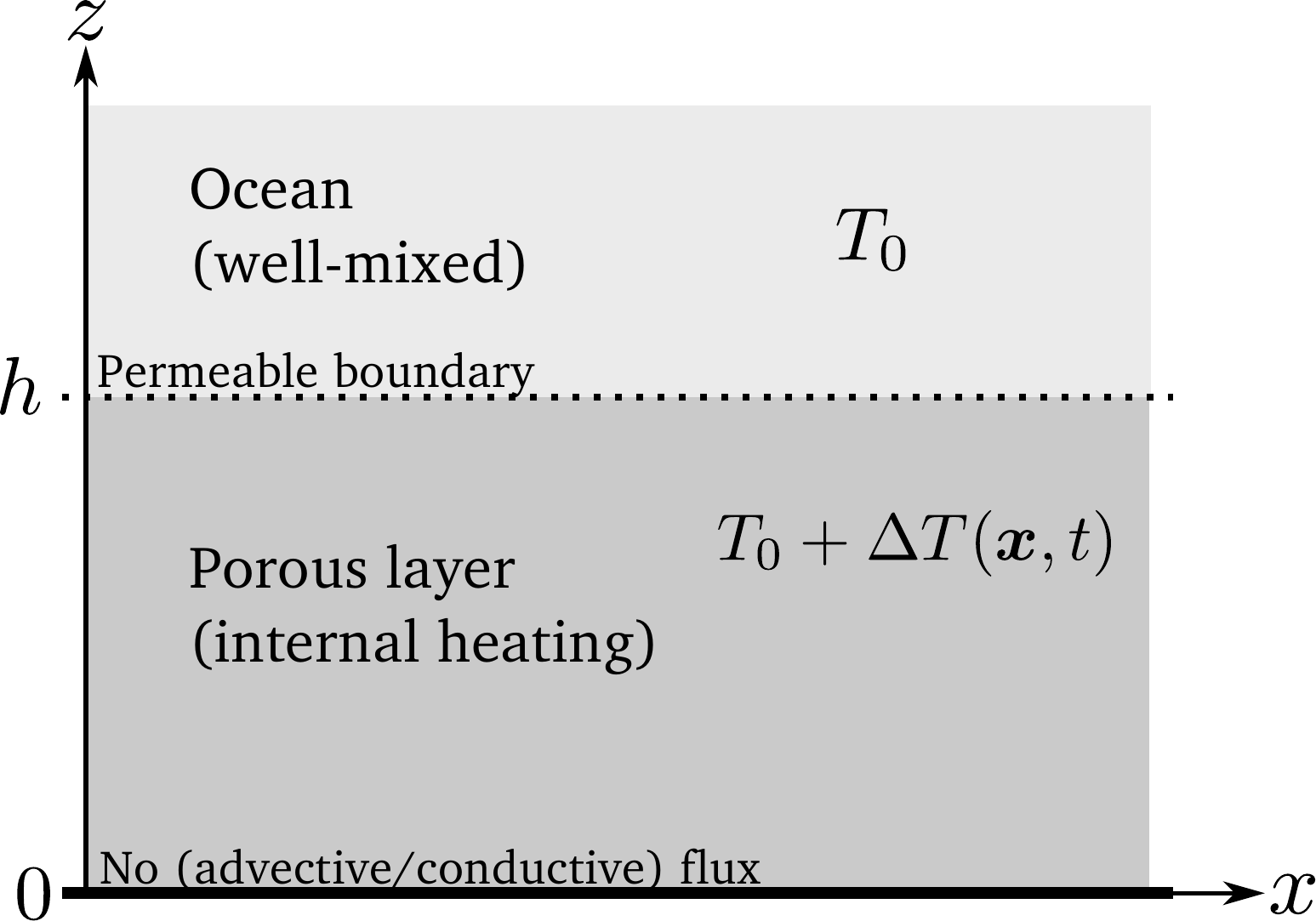}
\caption{An idealised two-dimensional model to describe porous convection inside core in interaction with the subsurface ocean \textcolor{RED}{in an icy moon of the type of Enceladus}. The bottom of the porous layer models the centre of the moon, there is no heat and mass flux at this height. Mass exchange between the ocean and the porous core are allowed with a free vertical velocity at the top. }
\label{fig:enceladus_analog_model}
\end{figure}

\textcolor{RED}{We consider an idealised model of tidally-driven convection inside icy moons of the type of Enceladus to focus on the effect of two fundamental ingredients: internal heating and an open top boundary. We thus make a series of simplifying approximations. 
First, rather than modelling the full fluid system, comprising the water-saturated core and the ocean, we consider only the core, and we parametrise the core--ocean interaction via boundary conditions that will be specified below.
Second, we consider a two-dimensional Cartesian model instead of modelling a full sphere. 
Third, we treat the gravitational field as constant in space, although it should increase away from the centre of the moon, and we consider either constant or horizontally varying internal heat generation, although it, too, should vary with depth. We treat all other parameters, including the permeability, as constants. 
We will return in section \ref{sec:application_to_enceladus} to consider and discuss the effect of some of these assumptions, as we apply our general findings and scalings to the case of Enceladus. 
}
%
%

%

%
We thus consider a two-dimensional porous core of (uniform) permeability $k$, which is saturated with water of viscosity $\mu$.
It lies beneath an ocean that we assume to be well mixed with a global temperature $T_0$ (see figure \ref{fig:enceladus_analog_model}).
The \textcolor{RED}{volume (or Darcy) flux} $\boldsymbol{U} = (U,W)$ inside the porous core is modelled by Darcy's law, 
\begin{equation}
\label{eq:Darcy}
\mbf{U} ~=~ \frac{k}{\mu} \left( -  \mbf{\nabla} P + \rho \mbf{g} \right)
\end{equation}
where $P$ is the pressure, $\rho$ is the density of water and $\mbf{g}$ is the gravity field, pointing in the $z$ direction.
\textcolor{RED}{Note the the volume flux $\mbf{U}$ is related to the fluid velocity $\mbf{U}_f$ by the porosity $\varphi$ of the matrix such that $\mbf{U} = \varphi \mbf{U}_f$. 
}
%
%
In addition to Darcy's law, the flow is assumed to be incompressible, so that \textcolor{RED}{the volume flux} must also satisfy a continuity equation,
\begin{equation}
\partial_x U + \partial_z W = 0~.
\end{equation}
Water motion inside the core is driven by buoyancy and temperature differences. 
We model the effects of temperature on density assuming linear expansion of the fluid with temperature under the Boussinesq approximation, such that $\rho = \rho_0 ( 1 - \alpha (T-T_0) ) $ where $\rho_0$ is a reference density and $\alpha$ the thermal expansion coefficient. 
Darcy's law may thus be written as,
\begin{equation}
\label{eq:Darcy_boussinesq}
\mbf{U} ~=~ \frac{k}{\mu} \left( -  \mbf{\nabla} P' + \rho_0 g \alpha \Theta \mbf{e}_z \right) 
\end{equation}
where $P' = P + \rho_0 g z$ and $\Theta \equiv T -T_0$.
Since the flow is driven by thermal anomalies $\Theta$, we must introduce an equation modelling the transport of heat inside the porous medium.
This is achieved using thermal energy conservation, in which a source term accounting for volume heat production is included \cite{nield_heat_2013,soucek_water_2014}:
\textcolor{RED}{
\begin{equation}
\label{eq:advection_diffusion}
\overline{\varphi}\,\partial_t \Theta + \mbf{U} \cdot \mbf{\nabla} \Theta ~=~ \kappa \nabla^2 \Theta + q
\end{equation}
}
with $\kappa$ the volume-averaged heat diffusivity inside the porous medium (\ie of both water and the porous matrix together), \textcolor{RED}{$\overline{\varphi}$ a modified porosity} and $q$ is the internal heat source term. 
\textcolor{RED}{Under the assumption of local thermal equilibrium between the fluid and the matrix, the modified porosity $\overline{\varphi}$ and the volume-averaged diffusivity are combinations of the porosity, $\varphi$, the heat capacity per unit of mass of the matrix and water, $c_m$ and $c_0$, and the density of the matrix and water, $\rho_m$ and $\rho_c$, such that 
\begin{equation}
\label{eq:modified_porosity}
 \overline{\varphi} = \frac{(1-\varphi) \rho_m c_m + \varphi \rho_0 c_0}{\rho_0 c_0} ~~~ \mbox{and} ~~~ \kappa = \frac{(1-\varphi) \lambda_m + \varphi \lambda_0}{\rho_0 c_0},
\end{equation}
\cite{nield_heat_2013,soucek_water_2014}, where the $\lambda_{m,0}$ are the heat conductivity of the matrix and water.
}
The source term $q$ is related to the volumetric heat production by tidal heating $Q_V$ via \textcolor{RED}{$q =Q_V/(\rho_0 c_0)$}.
%
In this paper, we consider two idealised limits: either $Q_V$ is constant or it is assumed to vary laterally (\ie in $x$) to model tidal heating inhomogeneities.

\subsection{Boundary conditions}

\textcolor{RED}{Throughout this work, we impose periodic boundary conditions in the horizontal direction. }
The bottom of the porous layer roughly corresponds to the core centre, and so we assume that there is no heat or mass flux crossing the bottom boundary, that is:
\begin{equation}
\label{eq:bottom_boundary}
\partial_z \Theta ( z=0) ~=~ 0 ~~~\mbox{and}~~~ W(z=0) = 0
\end{equation}
The top of the layer at $z=h$ is in contact with the ocean and must allow mass exchange between the core and the ocean.
This is achieved by imposing a purely vertical velocity at the top, \ie :
\begin{equation}
U (z= h) = 0~.
\end{equation}
The two layers are also thermally coupled, and we consider two possible boundary conditions for $\theta$ on the upper boundary.
One first natural choice is to impose the temperature (on the upper boundary) to be the temperature of the ocean, \ie, 
\begin{equation}
\label{eq:BC_1}
\Theta(z=h) = 0 
\end{equation}
However, in this case, the advective heat flux  driving hydrothermal activity $W \Theta (z= h)$ across the interface would vanish, which seems at odds with the idea that the water coming out the porous layer may drive a buoyant plume rising in the ocean. 
We could alternatively use another boundary condition where the temperature of water is left unchanged as it leaves the porous layer, while water enters with the imposed temperature of the ocean, that is,
\begin{equation}
\label{eq:BC_2}
\left\lbrace
\begin{array}{rl}
& \partial_z \Theta (z=h) = 1 ~~ \mbox{if}~W > 0 \\
& \Theta(z=h) = 0 ~~ \mbox{else}.
\end{array}
\right.
\end{equation}
%
%
%
Such a boundary condition is a standard parametrisation of core--ocean interactions \cite{rabinowicz_two_1998,
monnereau_is_2002,
cserepes_forms_2004,
choblet_powering_2017}.
The thermal boundary conditions (\ref{eq:BC_1}) and (\ref{eq:BC_2}) may be regarded as two end-members of the fully coupled problem of the core--ocean interaction. 
In the case of slow ascent in the porous medium, diffusion from the ocean inside the core causes the temperature inside the porous medium to drop in the top boundary vicinity.
Conversely, if the upwelling is fast, diffusion is not able to affect the temperature inside the ascending plume. 
As a side note, intermediary situations where $\partial_z \Theta (z= h) = - \beta$ with $\beta >0$ could also be considered.
Nevertheless, choosing between the two boundary conditions or parametrisation of $\beta$ would require a demanding study of the fully coupled system involving both the ocean and the porous core. 
We instead carry out \textcolor{RED}{two sets of} simulations using either boundary conditions (\ref{eq:BC_1}) and (\ref{eq:BC_2}). \textcolor{RED}{We will find that} the choice of boundary condition does not significantly affect the flow in the interior of the core. 

\subsection{Scaling the problem: dimensionless equations}\label{sec:scaling}
\label{sec:scaling_the_problem}

%
First, all considered lengths are normalised by the height of the porous layer $h$.
We must also define \textcolor{RED}{volume flux} and temperature scales, respectively denoted as $U^*$ and $\Theta^*$. 
Darcy's law (\ref{eq:Darcy_boussinesq}) gives a simple relation between these two scales,
\begin{equation}
\label{eq:temp_vel_scale}
U^* = \frac{k}{\mu} \rho_0 \alpha g \Theta^* ~.
\end{equation} 
Unlike in, say, Rayleigh-B\'enard set-up, the temperature scale $\Theta^*$ is not naturally imposed in the internally heated problem. 
We predict that in the non-linear regime, heat production and advection will be the dominant balance \textcolor{RED}{in (\ref{eq:advection_diffusion})}, leading to the following relation between the velocity and temperature scales, 
\begin{equation}
\label{eq:advection_production_balance}
U^* \Theta^* ~=~ h q~.
\end{equation}
Both scales then can be written as a function of physical parameters as follows: 
\begin{equation}
\label{eq:velocity_temperature_dimensional_scales}
U^{*2} = \displaystyle \frac{k}{\mu} \rho_0 \alpha g h q~~~~\mbox{and}~~~~
\Theta^* =\displaystyle \sqrt{\frac{\mu h q}{k \rho_0 \alpha g}}
\end{equation}
Given these scales, we find that the system is governed by only one dimensionless parameter, a Rayleigh number comparing the relative importance of advection and diffusion,
\begin{equation}
\label{eq:Rayleigh_number}
Ra ~\equiv~ \frac{h U^*}{\kappa} ~=~\left( 
\displaystyle
\frac{k \alpha g}{\kappa \nu } \frac{q h^2}{\kappa} h \right)^{1/2}.
\end{equation}
Note that other definitions have been considered for the Rayleigh number, depending in particular on the expected balance at play.
For instance, \citeA{buretta_convective_1976} choose \textcolor{RED}{volume flux} and temperature scales based on an advection and diffusion balance, rather than a balance between advection and heat production consider in (\ref{eq:advection_production_balance}), leading to a Rayleigh number $Ra_{bb} = Ra^2$.
Introducing the dimensionless temperature $\theta =\Theta/ \Theta^* $, \textcolor{RED}{volume flux} $\mbf{u} =  \mbf{U}/U^*$, the dimensionless governing equations for a porous layer with internal heating are:
\begin{equation}
\label{eq:dimensionless_equations}
\left\lbrace
\begin{array}{rl}
\bnabla \cdot \bu ~&=~ 0 \\
\bu ~&=~ -\bnabla p + \theta \mbf{e}_z \\
\partial_t \theta + \bu \cdot \bnabla \theta ~&=~ 
\displaystyle \frac{1}{Ra} \nabla^2 \theta + 1 
\end{array}
\right.
\end{equation}
where \textcolor{RED2}{time is normalised by $\overline{\varphi}h/U^*$ and pressure is rescaled by $\mu U^*/(h k)$.} 
The flow being incompressible and two-dimensional, it is convenient to introduce a stream function $\psi$ such that $\bu = \bnabla \times ( - \psi \mbf{e}_y)$.
The governing equations (\ref{eq:dimensionless_equations}) become
\begin{equation}
\label{eq:dimensionless_equations_psi}
\left\lbrace
\begin{array}{rl}
\nabla^2 \psi ~&=~ -\partial_x \theta \\
\partial_t \theta + \partial_z \psi \partial_x \theta - \partial_x \psi \partial_z \theta ~&=~ 
\displaystyle \frac{1}{Ra} \nabla^2 \theta + 1.
\end{array}
\right.
\end{equation}
Lastly the \textcolor{RED}{vertical} boundary conditions are
\begin{equation}
\label{eq:velocity_boundary_conditon}
w(z = 0) =
u (z = 1) = \partial_z \theta (z=0) = 0 ,
\end{equation}
and either 
\begin{align}
\label{eq:BC1_dimensionless}
&\mbox{BC 1:} ~~~ \theta(z=1) = 0~, ~~~\mbox{or} \\
&\mbox{BC 2:} ~~~\left\lbrace
\begin{array}{rl}
&\partial_z \theta (z=1) = 1  ~~~ \mbox{if}~w > 0,\\
\label{eq:BC2_dimensionless}
& \theta(z=1) = 0 ~~~\mbox{else},
\end{array}
\right.
\end{align}
for the temperature.
\textcolor{RED}{Note that the boundary conditions on the volume flux translate into
$\partial_z \psi (z = 1)  = \psi (z=0) = 0$. 
}
\textcolor{RED}{Lastly, the domain is periodic in the $x$ direction.}

\subsection{Numerical modelling}

We study this problem numerically with the code developed by \cite{hewitt_ultimate_2012}.
\textcolor{RED}{At each time step, Darcy's law is used to determine the stream function using Fourier transform in the horizontal direction and second order finite differences in the vertical direction.
The time evolution of the advection-diffusion equation is solved using an alternating direction implicit scheme \cite{press_numerical_1992}.
The diffusion term is discretised using standard second-order accurate finite differences and 
}
the use of two staggered grids for the stream function $\psi$ and the temperature field $\theta$ allows \textcolor{RED}{flux-conservative discretisation of the advection term.}
The finite difference \textcolor{RED}{in time is} second-order accurate as well. 
Anticipating strong gradients \textcolor{RED}{near the boundaries}, a vertical stretched grid is implemented to ensure the boundary layers are well resolved. 
The numerical \textcolor{RED}{discretisation} of equations (\ref{eq:dimensionless_equations_psi}) is tested in section \ref{sec:onset}.

\section{The onset of convection}
\label{sec:onset}

\begin{figure}
\centering
\includegraphics[width= \linewidth]{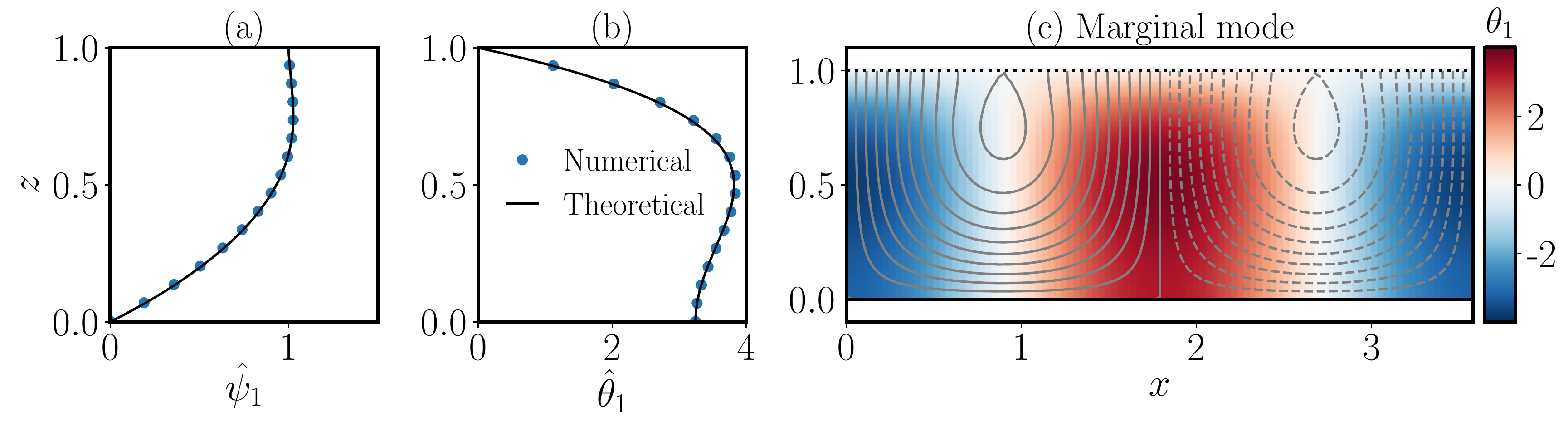}
\caption{(a) and (b): Vertical structure functions of the stream function $\hat{\psi}_{1}$ and $\hat{\theta}_{1}$ of the marginal mode obtained by solving the boundary value problem (\ref{eq:stability_f}) (black line) and extracted from a direct numerical simulation of the instability close to the threshold ($Ra - Ra_c \simeq 3 \times 10^{-2}$).
(c) Temperature field ($\theta_1$) and streamlines (iso-contours of $\psi_1$) of the unstable mode at the onset of convection.
} 
\label{fig:onset_structure}
\end{figure}

In this section, we investigate both theoretically and numerically the critical value of the Rayleigh number $Ra$ above which a convective instability develops. 
%
%
The steady, purely diffusive base $(\mbf{u} = 0, \theta_b)$ state on which the instability develops is
\begin{equation}
\label{eq:diffusive_base_state}
\theta_b(z) = \frac{Ra}{2} \left( 1 - z^2\right) 
\end{equation}
regardless of the upper thermal boundary condition.
We look for perturbations to the base state of the form \textcolor{RED}{\cite{drazin_introduction_2002} }:
\begin{equation}
\label{eq:stability_ansatz}
\psi = \psi_1 (\mbf{x}) e^{\sigma t} ~~~\mbox{and}~~~\theta = \theta_b + \theta_1 (\mbf{x}) e^{\sigma t}
\end{equation}
such that $\vert \psi_1 \vert ,~ \vert \theta_1\vert \ll \theta_b$.
The exponential terms allow to account for the existence of convective instability characterised by $\mathsf{Re}(\sigma) > 0$.
Using the ansatz (\ref{eq:stability_ansatz}), equations (\ref{eq:dimensionless_equations_psi}) to leading order in $\psi_1 $, $\theta_1$ yield the following single, fourth-order differential equation on the stream function:
\begin{equation}
\label{eq:stability_psi}
\nabla^4 \psi_1  = Ra \sigma \nabla^2 \psi_1 - z Ra^2 \partial_{xx} \psi_1.
\end{equation}
The invariance under translation along the $x$-axis allows further simplification by assuming that $\psi_1$ is a plane wave in $x$, that is $\psi_1 = \hat{\psi}_{1}(z) \exp(i k x)$.
Equation (\ref{eq:stability_psi}) with the plane wave assumption yields the following ordinary differential equation for the function $\hat{\psi}_{1}$:
\begin{equation}
\label{eq:stability_f}
\hat{\psi}_{1}^{''''} - (2 k^2 + Ra \sigma ) \hat{\psi}_{1}^{''} + (k^4 + Ra \sigma k^2 - z Ra^2 k^2) \hat{\psi}_{1} = 0 
\end{equation} 
where $\sigma$ is an unknown eigenvalue. 
We solve \textcolor{RED}{numerically the one-dimensional boundary value problem} (\ref{eq:stability_f}) using BC 1 in (\ref{eq:BC1_dimensionless}).
(In fact, for this onset problem, BC 2 (\ref{eq:BC2_dimensionless}) gives an ill-posed system.) 
We find the lowest value of the Rayleigh number for which $\sigma = 0$ to be $Ra= Ra_c \simeq 5.894$ at $k= k_c \simeq 1.751$.
Such a value for the critical Rayleigh number is close to the value $5.72$ found experimentally and theoretical by \citeA{buretta_convective_1976} in a system with closed boundary conditions.
The marginal mode and its vertical structure functions ($\hat{\psi}_{1}$ and $\hat{\theta}_{1}$) are shown in figure \ref{fig:onset_structure}.
The mode comprises a half-roll structure, with strong horizontal flow at the lower boundary and strong vertical flow at the upper boundary.
The temperature deviation is maximised roughly half-way up the roll.

We use this theoretical investigation of the onset of convection to benchmark the numerical code. 
Simulations were carried out at values of the Rayleigh number $Ra$ very close to the onset ($\vert Ra - Ra_c \vert \leq 10^{-1}$ typically).
The horizontal extent of the domain is chosen to match approximately twice the wave length of the marginal mode.
Computations were initiated with a small perturbation to the diffusive temperature profile (\ref{eq:diffusive_base_state}).
We observed an exponential growth or decay of the amplitude of the perturbation to the diffusive base state and found accurate reproduction of both the critical Rayleigh number and the growth or decay rate of the most unstable mode for nearby values of $Ra$. 
Figure \ref{fig:onset_structure} shows the excellent agreement between the theoretical and the computed vertical structure functions $\hat{\theta}_1$ and $\hat{\psi}_1$.

\section{Non-linear heat transport by convection}

\begin{table*}
\centering
\begin{tabular}{|l|l|l|l|}
\hline
Boundary conditions & Aspect ratio $L$ & Rayleigh number range &  Resolution ($n_x \times n_z$) \\ \hline
BC 1, 2 & 4 & 6-20 & $128\times 300$ \\
        &   & 20-100 & $256\times 300$ \\
		&   & 100-770 & $512\times 400$ \\
		&   & 550-2000 & $1024\times 500$ \\
		&   & 3000-10000 & $2048\times 500$ \\ \hline
BC  2   & 3 & 6-20 & $128\times 300$ \\
        &   & 20-100 & $256\times 300$ \\
		&   & 100-770 & $512\times 400$ \\
		&   & 550-2000 & $1024\times 500$ \\
		&   & 3000-10000 & $2048\times 500$ \\ \hline	
BC  2   & 8 & 6-20 & $256\times 300$ \\
        &   & 20-100 & $512\times 300$ \\
		&   & 100-770 & $1024\times 400$ \\
		&   & 550-3000 & $2048\times 500$ \\\hline
\end{tabular} 
\caption{Table of all the numerical simulations carried out indicating the nature of the boundary condition, the aspect ratio of the domain, the range of Rayleigh numbers and the associated horizontal ($n_x$) and vertical $(n_z)$ resolutions. Note that the resolution is increased close to the boundaries by the use of a stretched vertical grid.}
\label{tab:numerical_simulations_summary}
\end{table*}

In the following section, we investigate heat transport by convection for larger values of $Ra$. 
We first describe qualitatively the organisation of the flow as $Ra$ is increased.
We then show quantitatively that non-linear heat transport is dominated by advection, which constrains the typical size of hot plumes and thermal anomalies.
We use both thermal boundary conditions (\ref{eq:BC1_dimensionless}) and (\ref{eq:BC2_dimensionless}) to find that the difference between them is negligible for large enough values of $Ra$.

\subsection{Numerical process}
\label{sec:numerical_process}

Prior to delving into the results of the simulations, we explain how a typical numerical \textcolor{RED}{simulation} is carried out.
The simulations are initialised with  random noise at a certain Rayleigh number $Ra$. 
After the initial growth of the instability, the flow reaches a statistically steady state.
\textcolor{RED}{It is assessed by computing at each time step the mean of the maximum temperature since the start of the simulation: such a cumulative average converges towards a constant once the statistically steady state is reached.}
The simulation is terminated once the steady state has lasted for 300 time units.
%
%
The Rayleigh number is then switched to a new value, and the simulations is initiated with the last state of the previous one plus a small noise disturbance. 
A summary of all the numerical simulations that have been carried out is given in table \ref{tab:numerical_simulations_summary}.
%

\subsection{Flow structures and organisation}
\label{sec:flow_structure_organisation}

\begin{figure}
\centering
\includegraphics[width=\linewidth]{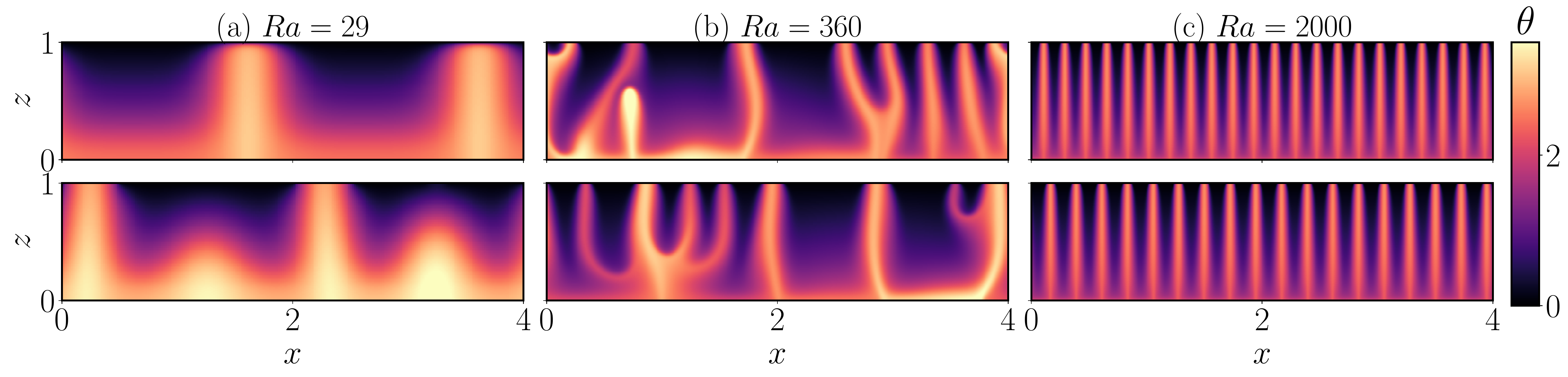}
\caption{Snapshots of the temperature field $\theta$ at $Ra = 29$, $360$ and $2000$ for boundary condition BC 1 (\textbf{top}) and BC 2 (\textbf{bottom}), \textcolor{RED}{taken once a statistically steady state is reached}. The flow  exhibits chaotic behaviour for the two lowest values of the Rayleigh number, and is steady at $Ra= 2000$.
Apart from the thin top boundary layer, both boundary conditions (\ref{eq:BC1_dimensionless}) and (\ref{eq:BC2_dimensionless}) overall produce the same flow.
The difference in the plume number at $Ra=2000$ between the two boundary conditions is rather due to the simultaneous stability of different modes. 
}
\label{fig:Ra100_field}
\end{figure}

To introduce the non-linear behaviour of the instability driven by internal heating, we first to illustrate typical flow patterns observed at different Rayleigh numbers. 
Figures \ref{fig:Ra100_field} displays typical snapshots of the temperature field.
At low Rayleigh number, \ie for $Ra_c \leq Ra <20$, the convection reaches a steady state with few plumes, be it for boundary condition BC 1 or BC 2 (see figure \ref{fig:Ra100_field}a). 
\textcolor{RED}{Similar} to the unstable mode at threshold, these plumes consist of half-rolls, although with steeper vertical gradients at the top boundary in the case of BC 1. 
For larger Rayleigh numbers (see \ref{fig:Ra100_field}b), the flow exhibits an unsteady chaotic behaviour where usually two modes with different number of plumes alternate, thus inducing chaotic merging and growth of plumes. 
This situation ceases for $Ra \simeq 600$, at least for an aspect ratio $L = 4$: higher values of the Rayleigh number give rise to steady solutions with a large number of narrow plumes (see figure \ref{fig:Ra100_field}c)
The only noticeable difference between the two boundary conditions is the existence of a thin thermal boundary layer when the top temperature is imposed (BC 1). 
Its thickness, of order $Ra^{-1}$, is set by a balance between vertical advection and diffusion. 
In addition, the high degree of similarity between the simulations carried out with different boundary condition suggests that the mixed boundary condition (BC 2) is reliable. 
Note that this is not the case below the threshold of the instability where flows that are highly sensitive to initial condition are observed. 
We therefore choose to use both boundary conditions in the study detailed hereafter, as long as $Ra > Ra_c$.
Lastly, \textcolor{RED}{note that in} these snapshots $\theta = \mathcal{O}(1)$, which confirms that the balance between advection and heat production drives the dynamics, a balance that was foreseen in section \ref{sec:scaling}.

\subsection{Advective heat transport}
\label{sec:advective_heat_transport}

The qualitative analysis of snapshots carried out in the preceding section indicates that advection dominates heat transport.
We propose in the following a quantitative analysis of the flow to support this assertion, in particular of the vertical temperature and heat flux profiles.
This analysis will allow us to compare internally heated porous convection with the more classical Rayleigh-B\'enard problem via the introduction of a generalised Nusselt number. 

\subsubsection{The mean temperature scale}

\begin{figure}
\centering
\includegraphics[width=\linewidth]{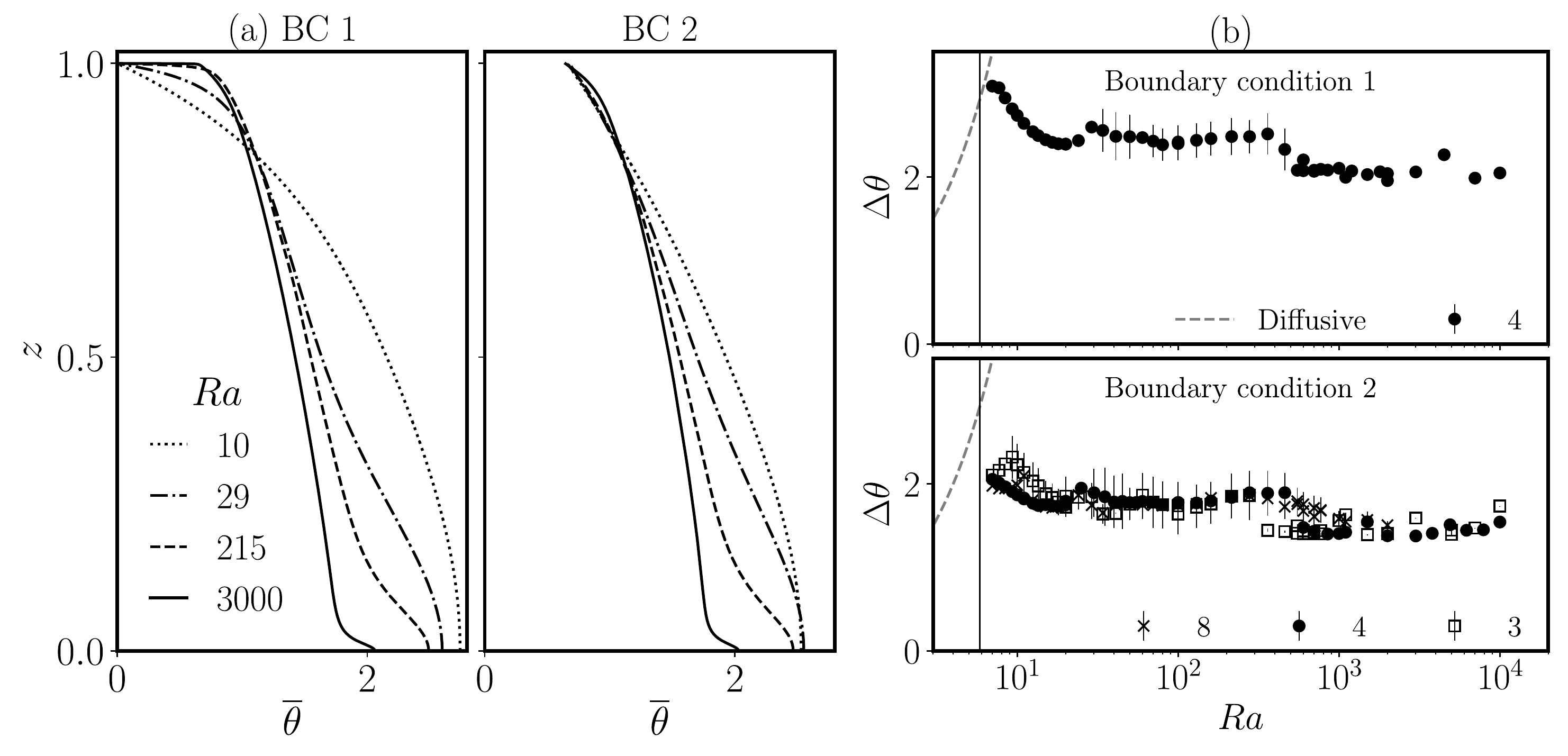}
\caption{\textbf{(a) }Averaged temperature profiles $\overline{\theta} (z)$ for different Rayleigh numbers and both boundary conditions and aspect ratio $L = 4$.
At large $Ra$, the profiles given by BC 1 and BC 2 are strikingly similar apart from the upper boundary layer for BC 1. 
\textcolor{RED}{
Note the emergence of a thermal boundary layer at $z=0$ where the heat that is produced locally is carried away through diffusion only.
Although it creates sharp variations, the bottom boundary condition $\overline{\theta}' (0) = 0$ remains satisfied even when rapid variations are observed at $Ra= 3000$. }
\textbf{(b)} Mean temperature difference between the bottom and the top of the porous layer $\Theta^*$ as a function of the Rayleigh number $Ra$, for all simulations carried out with BC 1 (top) and BC 2 (bottom).
\textcolor{RED}{
The errorbars correspond to the standard deviation over time of the average temperature difference. 
}
The diffusive temperature difference $\Delta \theta = Ra/2$ is shown for comparison (dashed line).
The vertical line marks the critical Rayleigh number $Ra_c$.
}
\label{fig:temperature_profile}
\label{fig:DetaT_Ra}
\end{figure}

It has been noted in the preceding section that the typical values of the temperature field remain of $O(1)$.
To better quantify this observation, we introduce a dimensionless temperature scale $\Delta \theta = \overline{\theta}(z=1) -\overline{\theta}(z=0)$, where the operation $\overline{\cdot}$ denotes horizontal and temporal average in the statistically steady state.
Typical profiles of the horizontally averaged temperature are shown in figure \ref{fig:temperature_profile}a. 
\textcolor{RED}{The average temperature is a decreasing function of height that converges towards an asymptotic profile at high Rayleigh number.
}
We note again the strong similarity between the two boundary conditions, especially at large Rayleigh numbers where they only differ by the presence of the top thermal boundary layer.
The temperature scale $\Delta \theta$ is also plotted in figure \ref{fig:DetaT_Ra} as a function of the Rayleigh number for all simulations.
We note it is well below the diffusive scaling $\Delta \theta \propto Ra$ even very close to the threshold of the instability. 
$\Delta \theta = O(1)$ is a signature of efficient transport and vertically mixing of the thermal energy by the convective flows.
In addition, we note a marked decrease of $\Delta \theta$ at $Ra \simeq 600$, which corresponds to the transition from the chaotic to the steady regime.
It indicates that the steady flow is even more efficient at transporting heat out of the system. 
Anomalous points may however be noticed; they are due to the locking of the simulation on a particular mode (\ie a flow with a certain number of plumes) that remains stable as the Rayleigh number is slightly increased.
\textcolor{RED}{We found that starting from a different initial condition at the same Rayleigh number can give steady states with a different number or plumes, which suggests that the past history of the system has some influence on its current state. }
%
%

\subsubsection{The advective flux}

\begin{figure}
\centering
\includegraphics[width=\linewidth]{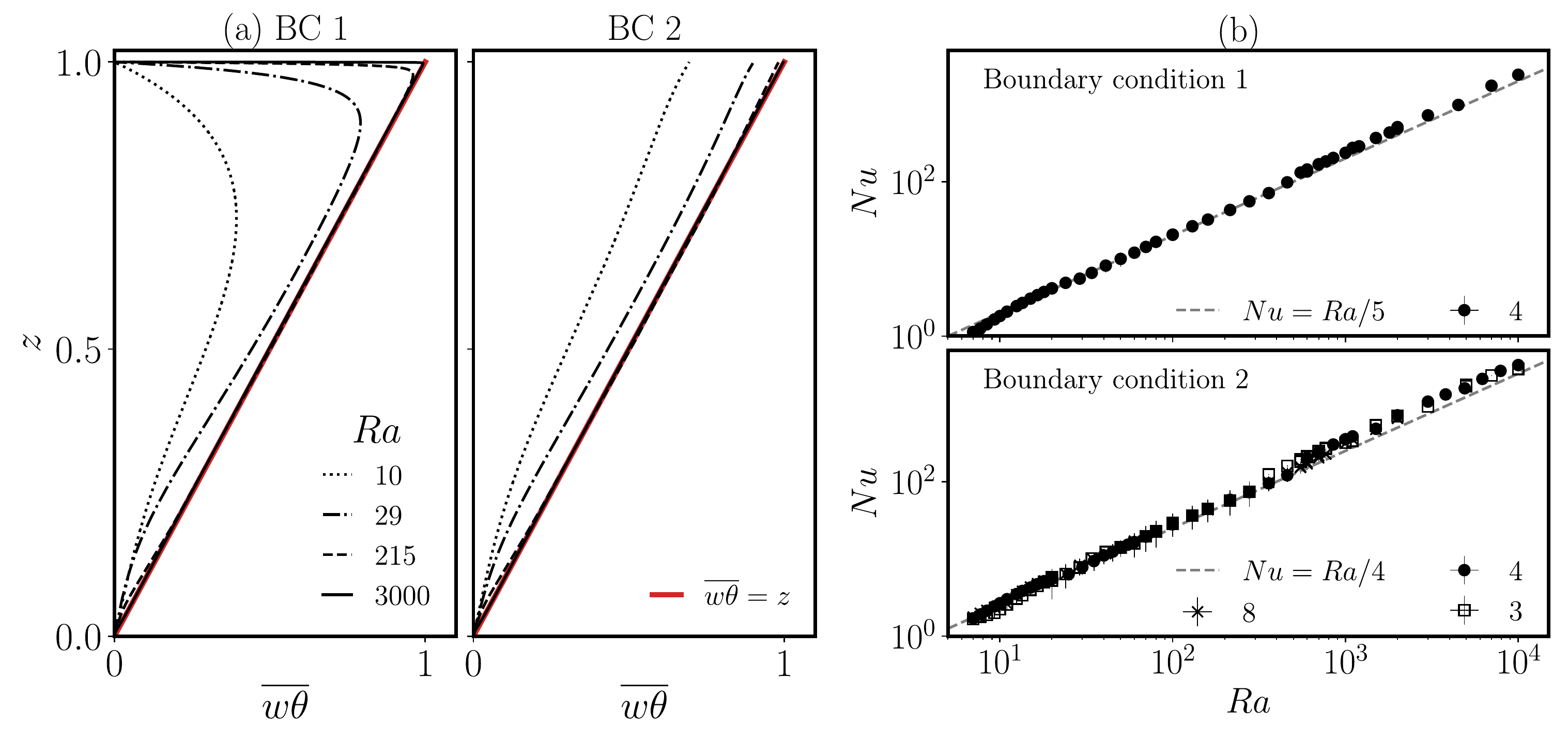}
\caption{\textbf{(a)} Vertical variations of the horizontally-averaged advective heat flux $\overline{w \theta} (z)$, for boundary condition BC 1 (left) and 2 (right) and aspect ratio $L = 4$. 
Again, we note the similarity between the two boundary condition in the bulk of the porous medium.
The asymptotic law $\overline{w \theta} = z$ (\ref{eq:asymptotic_advective}) is given for reference.
\textbf{(b)}  Plot of the Nusselt number $Nu$ as a function of the Rayleigh number $Ra$ for all simulations carried out with boundary conditions BC 1 \& 2 (top and bottom panels respectively).
\textcolor{RED}{The errorbars correspond to the standard deviation in time of the instantaneous Nusselt number $N(t)$ (see equation (\ref{eq:Nusselt_generalised})).}
 }
\label{fig:adv_prof_BC1}
\label{fig:N_Rdiag}
\end{figure}
To further quantify heat transport in the strongly non-linear regime, we consider here the vertical heat flux, defined as
\begin{equation}
\label{eq:heat_flux_conservation}
J = w \theta  - \frac{1}{Ra}  \frac{\partial \theta}{\partial z}~,
\end{equation}
which comprises an advective and a diffusive contribution.
Time-averaged thermal energy conservation (\ref{eq:dimensionless_equations}) prescribes a balance between vertical heat transport and volumetric heat production such that:
\begin{equation}
\label{eq:heat_flux_z}
\frac{\mathrm{d} \overline{J} }{\mathrm{d}z} = 1, ~~~\mbox{that is}~~~ \overline{J}(z) = z.
\end{equation}
In the asymptotic regime of high Rayleigh number, we expect that the heat produced is carried away by advection only, apart from the thermal boundary layer when it exists.
In the bulk of the porous medium, we thus expect
\begin{equation}
\label{eq:asymptotic_advective}
\overline{w\theta}(z) = z~.
\end{equation}
As can be noticed in figure \ref{fig:adv_prof_BC1}, the advective heat flux is well described by the asymptotic law (\ref{eq:asymptotic_advective}) even at Rayleigh numbers as low as $Ra = 29$. 
For BC 1, this agreement breaks down near the upper boundary where $\theta = 0$: the advection flux in the bulk is converted into a conductive heat flux over a boundary layer of depth $\mathcal{O}(Ra^{-1})$ which follows from equation (\ref{eq:heat_flux_conservation}).
%

\subsubsection{Nusselt number}

It is interesting to assess how efficient the convecting system is at transporting heat relative to purely diffusive transport.
It is quantified by a Nusselt number $N$ that provides a comparison between  the total heat flux (including advective and diffusive contributions) and the diffusive heat flux (see \textit{e.g.} \citeA{goluskin_family_2016}),
\begin{equation}
\label{eq:Nusselt_generalised}
N (t) \equiv \displaystyle \frac{\mean{ w \theta - Ra^{-1} \partial_z \theta}}{-Ra^{-1} \mean{\partial_z \theta}} =   \frac{Ra}{2 \Delta \theta}~,
\end{equation}
$\left \langle \cdot \right\rangle$ denoting volume average, and where we have used  that $\mean{w \theta - Ra^{-1} \partial_z \theta } = \mean{z} = 1/2$ and 
 $\mean{\partial_z \theta} = \Delta \theta $ from (\ref{eq:heat_flux_conservation}) and (\ref{eq:heat_flux_z}).
Then, we define the mean Nusselt number $Nu$ to be the long-time average of $N(t)$.
Note that we retrieve that the transport is purely diffusive at threshold, since, at $Ra = Ra_c$, $\Delta \theta = Ra/2$ so that $Nu = 1$.
%
%
%
Because $\Delta \theta = \mathcal{O}(1)$, we predict that, in the high Rayleigh number regime, $Nu \propto Ra$.
%
%
For both boundary conditions BC 1 \& 2, our simulations confirm this scaling down to $Ra \sim 20$ (see figure \ref{fig:N_Rdiag}b).
There is a slight enhancement of the efficiency of heat transport as steady states emerge in the non-linear regime of the instability around $Ra \sim 500$.
The same scaling between $Nu$ and $Ra$ is also found in the classical Rayleigh-B\'enard set-up in a porous medium \cite{otero_high-rayleigh-number_2004,hewitt_ultimate_2012,hewitt_high_2014}.
%
%

%

\subsection{Plume scales}

\begin{figure}
\centering
\includegraphics[width=\linewidth]{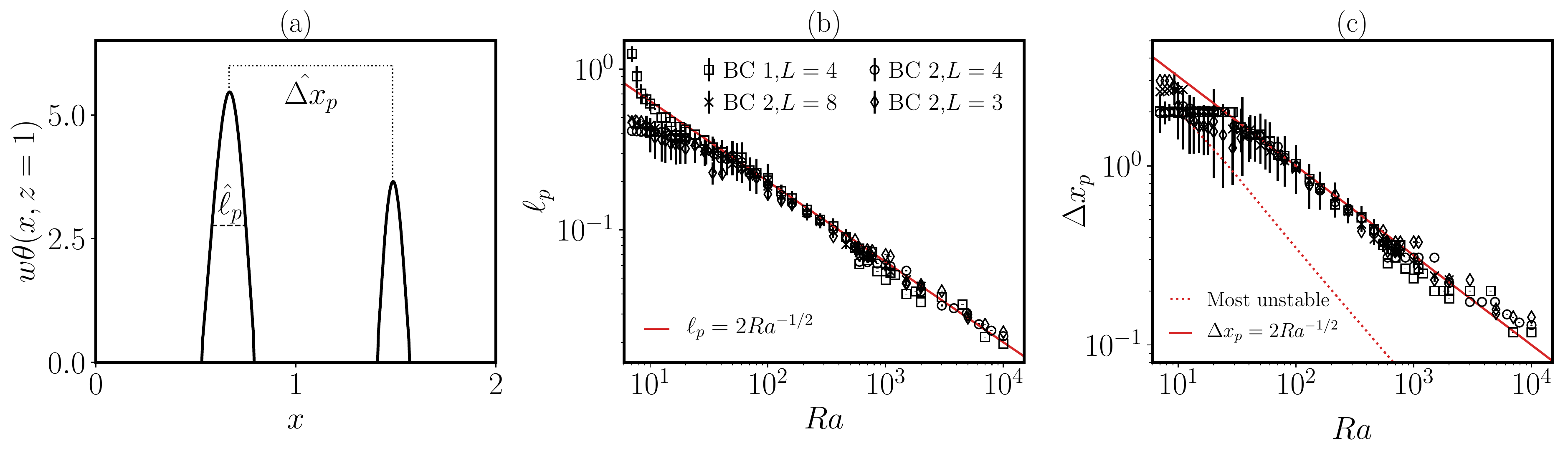}
\caption{
\textbf{(a)} vertical advective heat flux at the top boundary for $Ra= 200$ with boundary condition BC 2 and aspect ratio $L=4$. The plot focuses on two plumes and shows graphically the definition of the plume separation $ \hat{ \Delta x}_{p}$ and the plume width $\hat{\ell}_{p }$.
\textbf{(b, c)} mean plume size $ \ell_p$ and plume separation $\Delta x_p$, respectively, for all simulations with both boundary conditions BC 1 and BC 2.
Both quantities scale with the Rayleigh number as $Ra^{-1/2}$. 
The errorbars are determined by the standard deviation of the mean plume width and separation over a numerical run.
The red dashed line indicates the plume separation for the most unstable mode, which follows a $Ra^{-3/4}$ power law.
}
\label{fig:plume_scale}
\end{figure}

In section \ref{sec:flow_structure_organisation}, we observed that as the Rayleigh number $Ra$ is increased, the typical width of the plumes and their typical spacing decreases. 
%
We obtain a quantitative measure of the mean plume size $\ell_p$ and separation $\Delta x_p$ as a function of the Rayleigh number from the heat flux at the upper boundary. 
As shown in figure \ref{fig:plume_scale}a, plumes produce a series of heat-flux peaks. 
At each time step, we record the mean plume width $\hat{\ell}_{p}$ and plume separation distance $ \hat{\Delta x}_{p}(t)$ over all plumes, and we define $\ell_{p}$ and $\Delta x_p$ to be their long-time averages. 
Typical variability is given by the standard deviation of $\hat{\ell}_{p}$ and $\hat{\Delta x}_{p}$ over time. 
The result of this process is shown in figure \ref{fig:plume_scale}(b,c): both the plume width and separation exhibit the same scaling with the Rayleigh number, that is $\ell_p, \Delta x_p \propto Ra^{-1/2}$, even close to the threshold.
%
%
This power law can't be explained by linear theory, even at low $Ra$, as the mean separation between plumes does not coincide with the most unstable mode predicted by the linear stability analysis (see figure \ref{fig:plume_scale}c).
Instead, the typical scale of the plume is controlled by a balance between vertical advection, horizontal diffusion and heat production in (\ref{eq:dimensionless_equations}), 
that is,  
\begin{equation}
\label{eq:balance_plume_size}
w \partial_z \theta \sim Ra^{-1} \partial_{xx} \theta \sim 1.
\end{equation}
Given that temperature contrast remains $O(1)$, this balance demands both that the vertical velocity of the plume is $O(1)$ and that the typical lateral scale of the plume must be proportional to $Ra^{-1/2}$.

\subsection{Asymptotic plume solution}
\label{sec:asymptotic_plume}

Building on the scalings governing the typical plume size found numerically and theoretically, we derive here fully non-linear solutions of the equations (\ref{eq:dimensionless_equations}) in the asymptotic limit $Ra \rightarrow  \infty$. 
As explained below, the derivation of these equations allows us to understand the balance at play in the plume formation.
%

%
In the bulk of the porous medium, since the gradients are $O(Ra^{1/2})$ in the $x$ direction and $O(1)$ in the $z$ direction, the incompressibility condition $\p_x u + \p_z w$ imposes a scaling on the ratio between $u$ and $w$, that is $u/w = O(Ra^{-1/2})$.
We thus introduce the rescaled variables $\hat{x}$ and $\hat{u}$ such that:
\begin{equation}
x = Ra^{1/2} \hat{x} ~~~\mbox{and}~~~
u = Ra^{-1/2} \hat{u}~.
\end{equation}
With these rescaled variables, the incompressibility condition is
\begin{equation}
\partial_{\hat{x}} \hat{u} + \partial_z w = 0
\end{equation}
Taking the curl of Darcy's law in (\ref{eq:dimensionless_equations}) yields 
\begin{equation}
\label{eq:curl_Darcy_rescaled}
\partial_{\hat{x}} w  =  \partial_{\hat{x}} \theta + O(Ra^{-1})~.
\end{equation}
\textcolor{RED}{Hence, to leading order in $Ra$, $\theta-w$ is a function of $z$ only.
Because $\overline{w} = 0$, we infer that $\theta (\hat{x},z) = w(\hat{x},z) + \overline{\theta}(z)$ .}
Thus, Darcy's law compels the temperature and the vertical velocity to have the same horizontal variance.
Lastly, the advection-diffusion equation in (\ref{eq:dimensionless_equations}) with rescaled variables is
\begin{equation}
\label{eq:rescaled_advection_diffusion}
\partial_t \theta +\hat{u} \partial_{\hat{x}} \theta + w \partial_z  \theta =  \partial_{\hat{x}\hat{x}} \theta + 1~.
\end{equation}
where all terms appear to be of the same order.
Building on our numerical results, we seek steady solutions that are periodic in the $x$ direction.
We introduce an ansatz for the flow that is the lowest order truncation of a Fourier series, that is, we assume the velocity field to have the following form,
\begin{equation}
\label{eq:velocity_ansatz}
\left\lbrace
\begin{array}{rl}
\hat{u} ~&=~ \hat{u}_0 (z) \sin(\hat{k} \hat{x}) \\
w ~&=~ w_0(z) \cos(\hat{k} \hat{x})  ~,
\end{array}
\right.
\end{equation}
which has no mean mass flux in either vertical or horizontal direction.
%
%
According to the rescaled Darcy's law (\ref{eq:curl_Darcy_rescaled}), the temperature becomes
\begin{equation}
\label{eq:temperature_ansatz}
\theta = w + \overline{\theta}(z) =  w_0(z) \cos(\hat{k} \hat{x}) + \overline{\theta}(z)~.
\end{equation}
For the flow (\ref{eq:velocity_ansatz}) to satisfy the incompressibility condition, the following relation is required:
\begin{equation}
\label{eq:continuity_ansatz}
w_0' = - \hat{k} \hat{u}_0~.
\end{equation}
To determine the functions $w_0$ and $\overline{\theta}$, we use the advection-diffusion equation (\ref{eq:rescaled_advection_diffusion}) \textcolor{RED}{which becomes}
\textcolor{RED}{
\begin{equation}
\label{eq:adv_diff_harmonics}
w_0 w_0' \frac{1-\cos(2 \hat{k}\hat{x} ) }{2} +  w_0 w_0' \frac{1+\cos(2 \hat{k}\hat{x} ) }{2} + w_0 \overline{\theta}' \cos(\hat{k}\hat{x} ) = - \hat{k}^2 w_0 \cos(\hat{k} \hat{x}) + 1
\end{equation}
where the incompressibility condition (\ref{eq:continuity_ansatz}) and Darcy's law (\ref{eq:curl_Darcy_rescaled}) have been used.}
\textcolor{RED}{This equation contains mean and $\hat{k}$ harmonic terms that must be balanced, respectively.}
%
%
The mean terms 
simply yields a balance between vertical heat advection and heat production, that is:
\begin{equation}
\ddroit{w_0^2}{z} = 2 ~~~\mbox{\ie} w_0 = \sqrt{2 z}~.
\end{equation}
The harmonic $\hat{k}$ terms correspond to a balance between horizontal diffusion and the vertical advection of the average thermal energy (or temperature) profile,
\begin{equation}
w_0 \overline{\theta}' = -\hat{k}^2 w_0~~~ \mbox{\ie} \overline{\theta}(z) = \theta_0 - \hat{k}^2 z~.
\end{equation}
As noted in the preceding section, such a balance is responsible for setting the $O(Ra^{1/2})$ horizontal gradients.
%
%

%
We have, therefore, constructed a fully non-linear solution that is exact in the asymptotic limit $Ra \rightarrow \infty$, and is given by 
\begin{equation}
\label{eq:nonlinear_solution}
\left\lbrace
\begin{array}{rl}
u &= -  \displaystyle \frac{Ra^{-1/2}}{\hat{k} \sqrt{2z}} \sin(Ra^{1/2} \hat{k} x) \\[1em] 
w &=  \sqrt{2z} \cos(Ra^{1/2} \hat{k} x) \\[1em]
\theta &= \theta_0 - \hat{k}^2 z + \sqrt{2z} \cos(Ra^{1/2} \hat{k} x)
\end{array}
\right.
\end{equation}
where $\hat{k}$ and $\theta_0$ are $O(1)$ but \textit{a priori} unknown.
Note that this solution only satisfies one boundary condition: the absence of mass flux at the bottom of the porous layer.
The remaining boundary conditions, be it the absence of bottom heat flux, the purely vertical velocity at the top, or either of the thermal boundary condition BC 1 or BC 2, are all unmatched with the solution.
%

\begin{figure}
\centering
\includegraphics[height =0.33\linewidth]{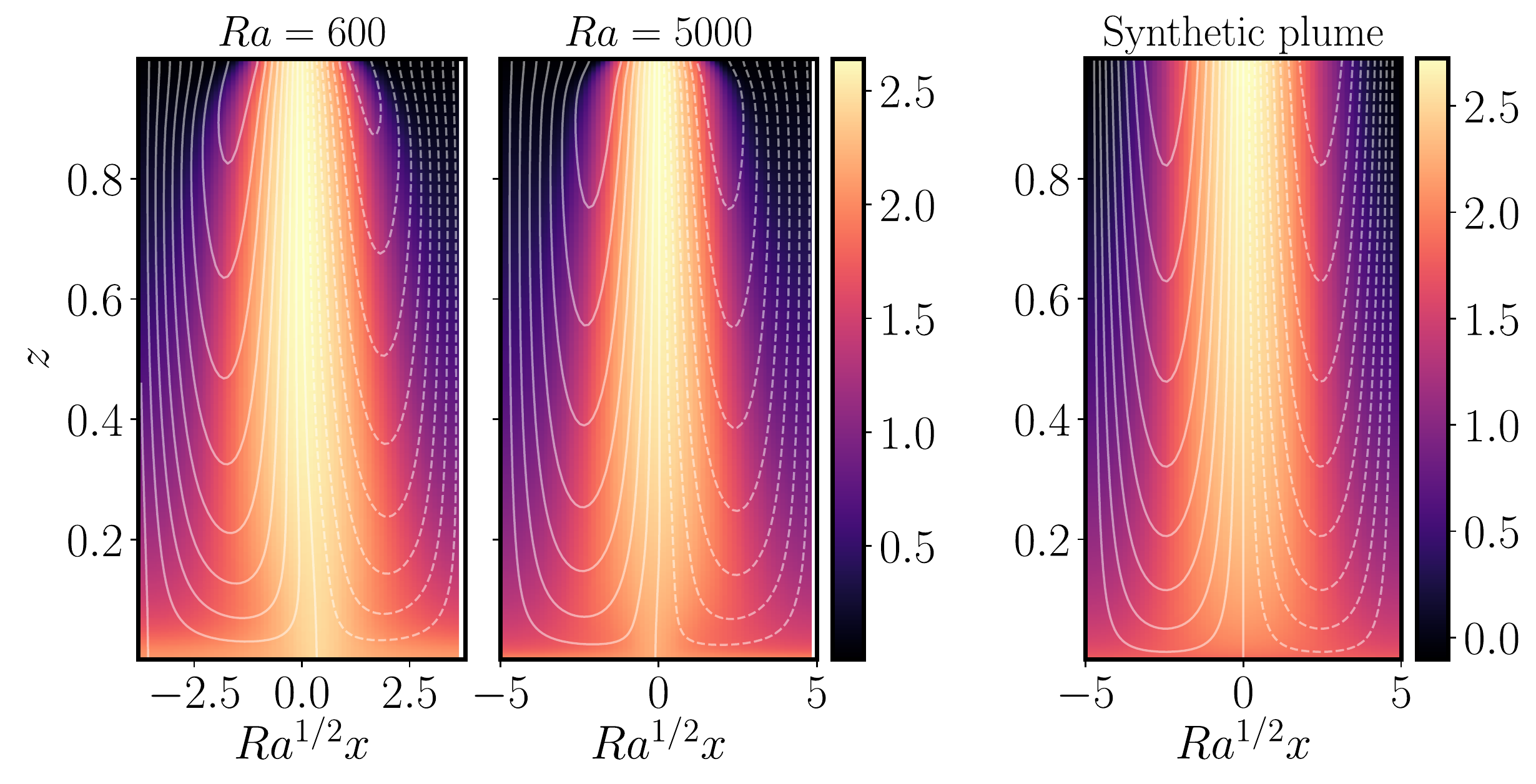}
\caption{\textbf{Left:} single plume isolated in the regime where the flow is steady and periodic in $x$, with boundary condition BC 1. \textbf{Right:} synthetic field for a single plume obtained from the solution (\ref{eq:nonlinear_solution}). The rescaled wave number $\hat{k}$ is chosen to match the $Ra= 5000$ case, and its value is around $0.64$. The integration constant $\theta_0$ is chosen around 1.8 to roughly match the bottom temperature profiles observed in \ref{fig:temperature_profile}. Note that although the streamlines do not seem to be vertical at the top boundary, a zoom shows that they ultimately bend to match verticality very close to the top boundary. } 
\label{fig:plume_studies}
\end{figure}

%
Figure \ref{fig:plume_studies} provides a comparison between plumes extracted from the simulations at two different Rayleigh numbers with a synthetic plume corresponding to the solution (\ref{eq:nonlinear_solution}).  
The overall qualitative behaviour of the two fields are the same, although the theoretical solution does not capture the shrinking of the plumes close the the top boundary, because it does not satisfy the boundary condition there.

\begin{figure}
\centering
\includegraphics[width=\linewidth]{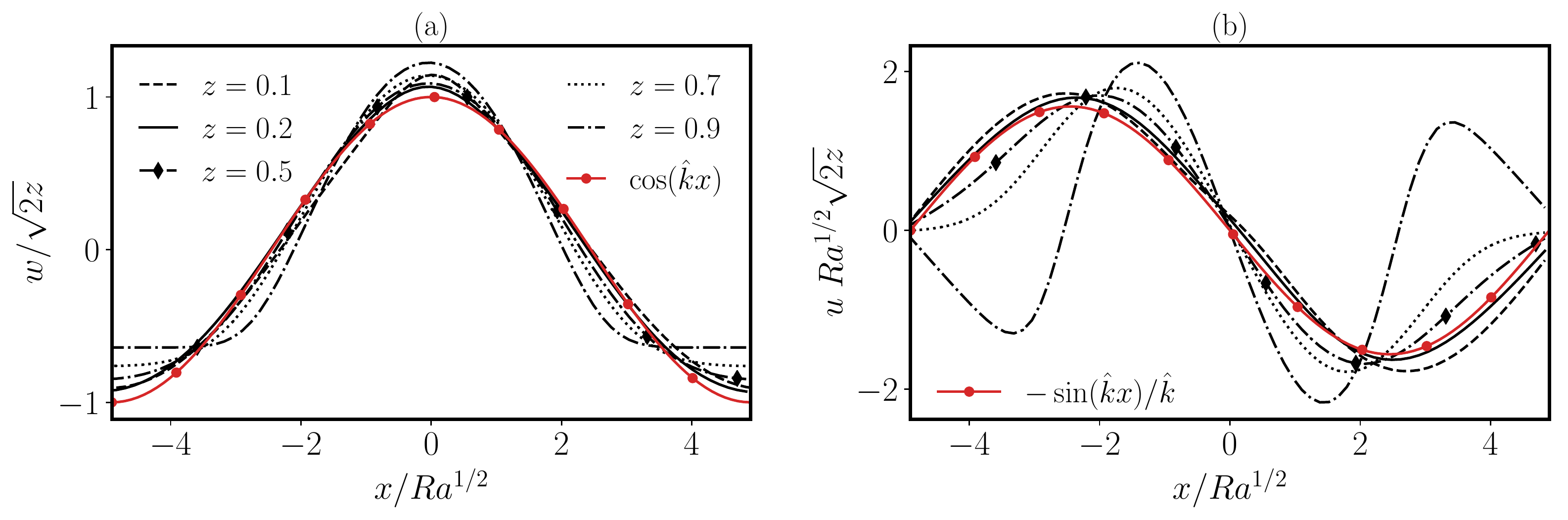}
\caption{Horizontal profiles at different heights $z$  of the scaled vertical \textbf{(a)} and horizontal~\textbf{(b)} velocity across a plume. The amplitudes are normalised accordingly to the non-linear solution (\ref{eq:nonlinear_solution}). The expected structure is shown in red and the only fitting parameter is the rescaled wave number $\hat{k} \simeq 0.64$. The Rayleigh number is $Ra = 5000$ and the top boundary condition is BC 2.}
\label{fig:plume_profiles_fit}
\end{figure}

To draw a more quantitative comparison between the non-linear solution and the flow in one plume, we plot in figure \ref{fig:plume_profiles_fit} several horizontal cuts at different heights of the vertical and horizontal velocities.
We find that in the bulk, the theoretical solution adequately describes the amplitude of the velocity variations.
However, the model becomes inaccurate near the upper boundary where, as noted above, it does not satisfy  the correct boundary conditions.
In fact, this issue seems to lead to other inaccuracies in the model: it predicts a linear decrease in the mean temperature $\overline{\theta}$, whereas the numerical simulations show a more complex dependence on $z$ (figure \ref{fig:temperature_profile}).
%
%
%

\subsection{Conclusion for non-linear heat transport}

Throughout this section, we have detailed the properties of heat transport by convection in strongly non-linear regimes.
Based on several arguments, including temperature scale, heat flux and Nusselt number measurement, we have confirmed that heat transport is dominated by advection in the bulk of the porous medium.
We have carried out simulations with two different top boundary conditions that are thought to be relevant to geophysical context: one where the top boundary temperature is imposed, and another where advective heat flux is conserved in upwellings and temperature is imposed in downwellings. 
We have confirmed that the two boundary conditions produce the same bulk flows.
Lastly, we have shown that the typical plume size follows a $Ra^{-1/2}$ power law, which is due to a balance between horizontal diffusion and vertical advection of heat.
It is interesting to note that internally-heated and Rayleigh-B\'enard convection are different regarding the typical plume scale: in the asymptotic regimes of large $Ra$, \citeA{hewitt_ultimate_2012} found that $\ell_p$ scaled like $Ra^{-0.4}$, which they later suggested was a result of the stability of the plumes \cite{hewitt_stability_2017-2}. 
%


\section{Accounting for the large scale modulation of tidal heating}
\label{sec:heating_large_scale_modulation}

\subsection{A simple model}

In this section, we briefly explore how the large-scale variations of tidal heating affect heat transport in internally heated porous media. 
This is important for the case of icy satellites such as Enceladus, for which heterogeneity of tidal heating have been shown to induce focusing of  the heat flux where heating is the most intense \cite{choblet_powering_2017}.
We consider here a domain with aspect ratio $L = 4$ for which the volume production of heat $q$ takes the form
\begin{equation}
\label{eq:heating_modulation}
q(x) = 1 - \Delta q \cos\left(\frac{2 \pi}{L} x\right)
\end{equation}
which is such that the mean heat production is unchanged compared the homogeneous case and the maximum heat production is located at the centre of the domain. 
In the following, we only illustrate heat modulation with $\Delta q = 0.5$ in the case of the boundary condition BC 2. 
$\Delta q = 0.5$ is a good proxy for tidal heating which bears latitudinal and longitudinal variations by about a factor 2 between minima and maxima.

\subsection{Large scale flow and pulsatility}

\begin{figure}
\includegraphics[width=\linewidth]{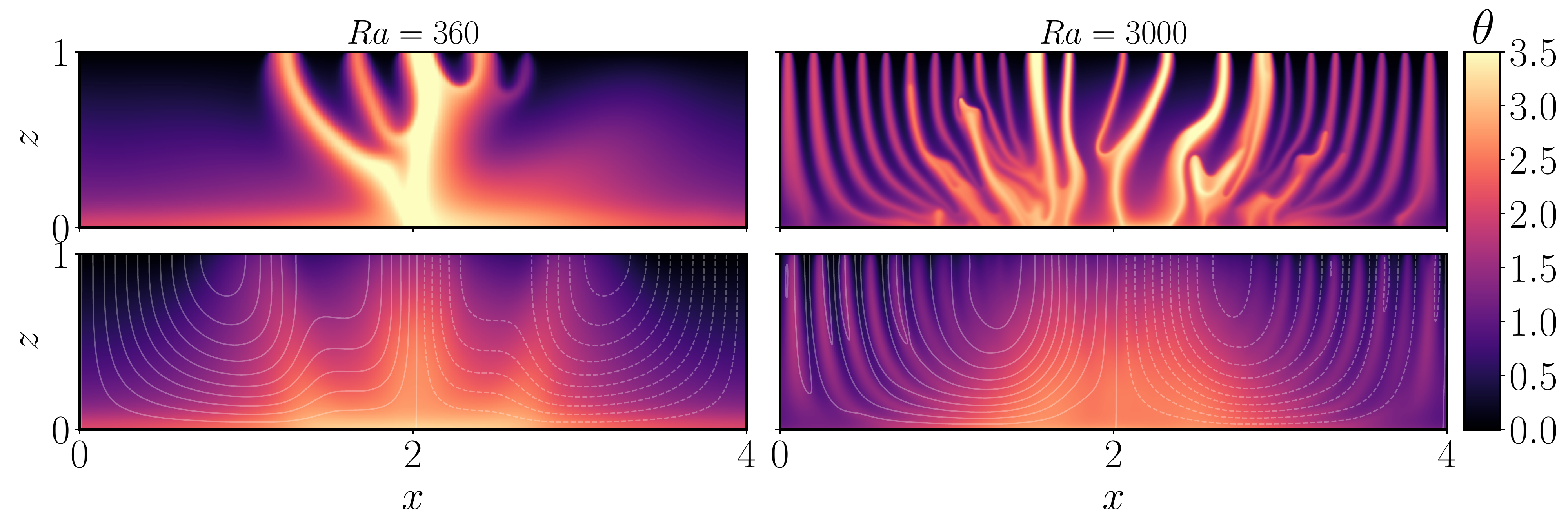}
\caption{Snapshot of the temperature field for the heterogeneous heating case (\textbf{top}) and the time-averaged temperature and flow streamlines averaged (\textbf{bottom}).}
\label{fig:heterogeneous_heating_snapshots}
\end{figure}

\textcolor{RED}{The large-scale modulation of internal heating} leads to the emergence of several striking features.
The first one is the attraction of plumes towards the centre where the heating is the most intense. 
Although plumes may exist in the whole interior of the domain, they merge towards the centre, which results in a higher temperature region with larger heat flux anomaly, as illustrated in figure \ref{fig:heterogeneous_heating_snapshots}. 
Plume merging towards the centre is associated with a large-scale mean flow that is also shown in figure \ref{fig:heterogeneous_heating_snapshots}. 
\textcolor{RED}{
Note that at high Rayleigh number ($Ra= 3000$ in the snapshot of figure \ref{fig:heterogeneous_heating_snapshots}), small-scale plumes persist in the time-averaged flow.
Despite strong variability where heat production is maximal, some plumes remain locked in the areas where heat production is minimal, a feature that is reminiscent of the steady plumes of the homogeneous case. 
}

\begin{figure}
\centering
\includegraphics[width=\linewidth]{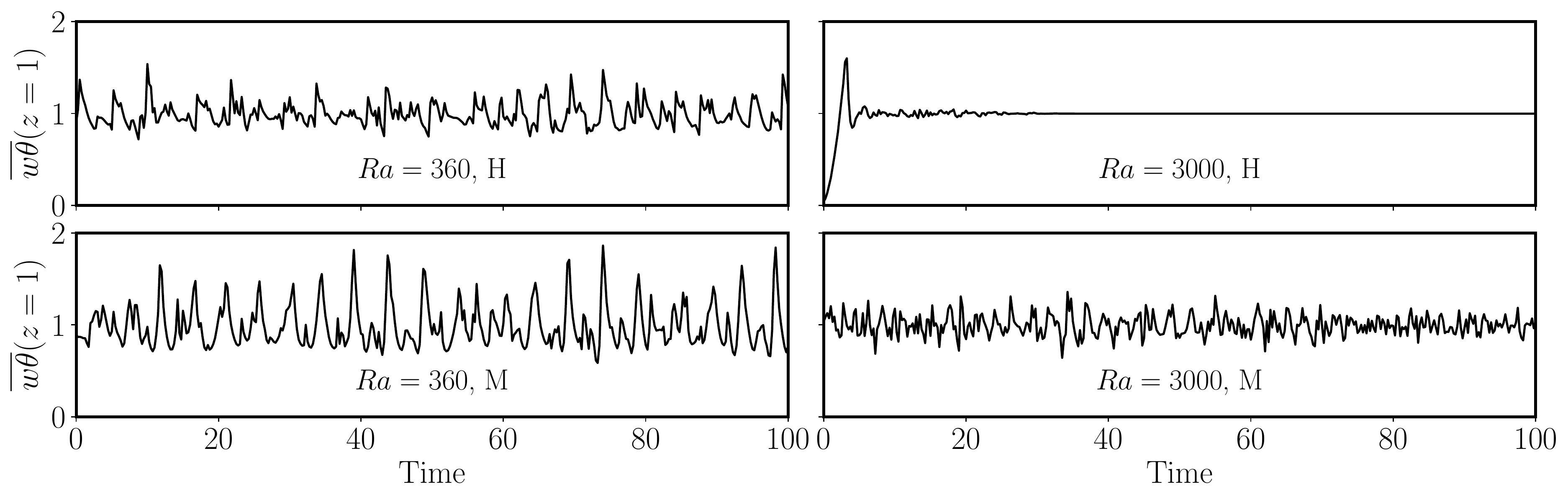}
\caption{Horizontally averaged advective heat flux at the top boundary at $Ra=360$ (\textbf{left}) and $Ra =3000$ (\textbf{right}), with comparison between homogeneous (\textbf{top}, H) and heterogeneous (\textbf{bottom}, M).  }
\label{fig:flux_time_mod}
\end{figure}
Advection of the plumes towards the largest internal heating region and the subsequent plume merging leads to pulsatility in the advective heat flux, as shown in figure \ref{fig:flux_time_mod}. 
At intermediate Rayleigh number ($Ra = 360$), the flux is intermittent for homogeneous heating but it exhibits a quasi-periodic behaviour for a modulated heating.
The typical period  is of order one, \ie it takes place over a convective time scale, and corresponds to the time needed for plume formation, advection towards the centre and merging. 
The effects of heterogeneous heating are even more striking at high $Ra$: the steady state observed in the homogeneous case is replaced by quick oscillations of the heat flux (see figure \ref{fig:flux_time_mod}).
They are due to the many plumes observed in the centre of the domain reaching the top boundary non synchronously (see figure \ref{fig:heterogeneous_heating_snapshots}).


\subsection{Similarities with the homogeneous-heating case}

\begin{figure}
\centering
\includegraphics[width=\linewidth]{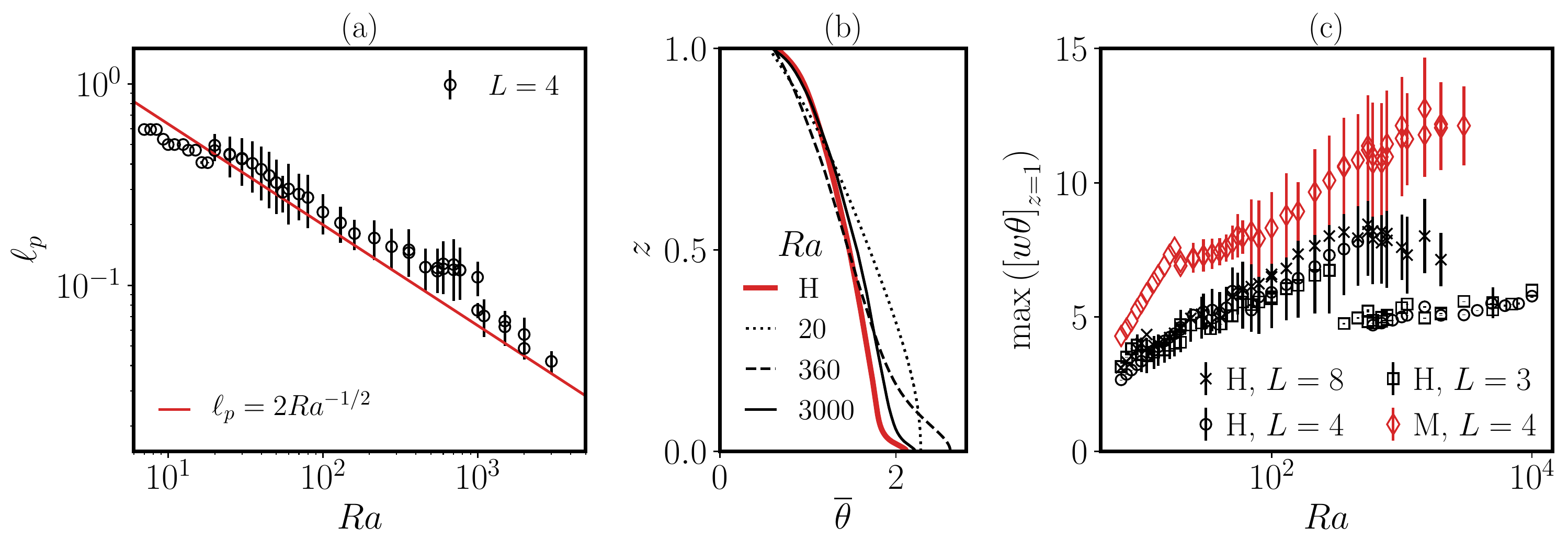}
\caption{\textbf{(a)} Typical plume size in the heterogeneous heating case, and comparison with the same law $\ell_p \propto Ra^{-1/2}$ as in figure \ref{fig:plume_scale}.
The errorbars are determined in the same way as in figure \ref{fig:plume_scale}.
\textbf{(b)} Mean temperature profile for heterogeneous heating at several Rayleigh numbers; the red line correspond to the temperature profile in the homogeneous case in the high $Ra$ regime.
\textcolor{RED}{\textbf{(c)} Maximum value of the non-dimensional advective heat flux at the top of the porous layer determined from the simulations with BC 2. The error bar accounts for the standard deviation of the maximum value over the course of a simulation. Both homogeneous ($\Delta q = 0$, labelled H) and heterogeneous ($\Delta q = 0.5$, labelled M) are shown.}
}
\label{fig:similarities_hetero}
\end{figure}

Despite the existence of a mean flow and the pulsatile behaviour detailed in the preceding section, convection with heterogeneous internal heating bears many similarities with the homogeneous case. 
As already noticed earlier, small scale plumes are still present in the flow, and their typical width remains proportional to $Ra^{-1/2}$ (see figure \ref{fig:similarities_hetero}) but with increased temporal and spatial variability.
This means that the balance between horizontal diffusion, heat production and vertical advection is still at play to determine the single plume dynamics. 
Moreover, even if lateral variations of the mean temperature are obvious in figure \ref{fig:heterogeneous_heating_snapshots}, the horizontally averaged temperature follows a trend that is very close to the homogeneous case, as shown in \ref{fig:similarities_hetero}. 
This observation suggests that the spatial form of heating is not particularly important for the mean dynamics and scaling laws governing heat transport in an internally heated porous medium.

\subsection{Hydrothermal velocity driven in the ocean}

\textcolor{RED}{
To conclude this theoretical analysis of internally heated porous convection, we derive a scaling law for the hydrothermal velocity driven by the hot plumes in the porous medium as they cross the upper boundary into the ocean above. 
}
To evaluate the typical velocity of the buoyant hot water coming out of the core at the bottom of the ocean we must first determine the buoyancy flux associated with the porous plumes. 
\textcolor{RED}{
The first question that arises concerns the transposition of the plumes observed in the present model to a three-dimensional geometry, as upwellings may take the form of isolated plumes or ascending sheets.
Although there is not a clear theoretical argument in favour of one or the other, several studies point towards the formation of sheets.
\citeA{monnereau_is_2002} has shown in viscous convection that the opening of the top boundary leads to a transition from plumes to sheets.
In porous flows, sheet-like convection is observed in simulations of hydrothermal flows \cite{rabinowicz_two_1998}, and, in particular, in the model of \citeA{choblet_powering_2017} for the core of Enceladus. 
Therefore, we assume in the following that upwelling in the porous medium takes the form of sheets.
As in the two-dimensional case, their typical extent is $Ra^{-1/2}$ because it remains set by the balance between vertical advection, horizontal diffusion and heat production given in equation (\ref{eq:balance_plume_size}). 
At the bottom of the ocean, the sheets produce a line source of buoyancy flux $B$ which drives hydrothermal velocities $U_h$ of order $B^{1/3}$.  \cite{morton_turbulent_1956,woods_turbulent_2010}.
}
The dimensional buoyancy flux is given by \cite{woods_turbulent_2010}
\begin{equation}
B ~=~ \int_{\mbox{upwelling}} \alpha g \left.(\Theta W)\right\vert_{z=h} \, \mathrm{d} x,
\end{equation}
where the one-dimensional integral is computed across an upwelling zone of typical extent $\ell_p \propto Ra^{-1/2}$. 
As $\Theta$ and $W$ are proportional to $Ra$, $B$ scales like $Ra^{3/2}$, or more explicitly,
\begin{equation}
\label{eq:buoyancy_2d}
B \simeq \frac{\kappa^2 \nu}{k h} ~ Ra^{3/2} ~ \left.( w \theta)\right\vert_{z=h}~~~~ \mbox{and}~~ U_h = \left(\frac{\kappa^2 \nu}{k h} \left.( w \theta)\right\vert_{z=h}  \right)^{1/3}  Ra^{1/2}
\end{equation}
\textcolor{RED}{where we have used the velocity and temperature scales defined in equation (\ref{eq:velocity_temperature_dimensional_scales}).}
Focusing of the heat flux in narrow upwelling zones leads to enhanced values of $\left.( w \theta)\right\vert_{z=h}$, as shown in figure \ref{fig:similarities_hetero}c. 
In the case of heterogeneous heating, focusing increases the heat flux at the bottom of the ocean by about a factor of $10$ over the range of Rayleigh numbers considered here.
%
%

%

\section{Application of the idealised study to the case of Enceladus}
\label{sec:application_to_enceladus}

\textcolor{RED}{
\subsection{From a two-dimensional model to a planetary core.}
}
\label{sec:ingredients_discussion}

\begin{figure}
\centering
\includegraphics[width=0.9\linewidth]{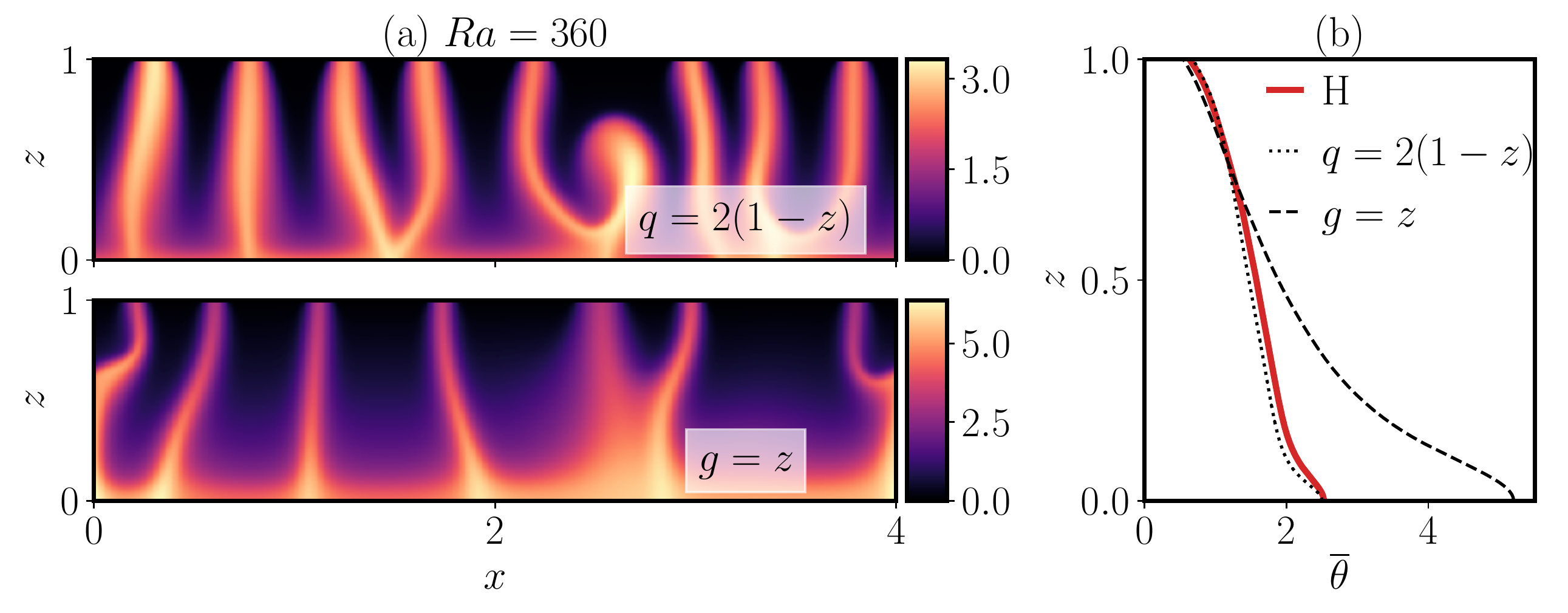}
\caption{ \textbf{(a)} Snapshots of the temperature field taken in the statistically steady state of simulations with $Ra= 360$. In the upper one, the dimensionless heat production rate is a decreasing function of height $q = 2(1-z)$ whereas in the lower one, the dimensionless gravity is a linear function of height $g = 2z$. \textbf{(b)} Superposition of the average temperature profiles at $Ra = 360$ for the homogeneous heating (H), height-dependent heating and height-dependent gravity cases.   }
\label{fig:Ra360_gz_snapshot} 
\end{figure}

\textcolor{RED}{
To carry out our idealised study of internally heated porous convection, we have discarded many ingredients that will be important for Enceladus, as stated prior to the introduction of the model (section \ref{sec:model_introduction}). 
Before applying our results, we review these approximations and evaluate how they may affect the conclusions drawn from the two-dimensional Cartesian model. 
}

\subsubsection*{Sphericity}
\textcolor{RED}{
First, the definition of the velocity, temperature and time scales as well as the Rayleigh number defined in section \ref{sec:scaling_the_problem} rely entirely on dimensional analysis and are thus insensitive to the geometry. 
The dimensionless equations (\ref{eq:dimensionless_equations}) thus take the same expression in any geometry. 
The critical Rayleigh number above which convection takes place will, presumably, be modified by the geometry, although \cite{choblet_powering_2017} found convective flows down to $Ra \simeq 8$ in simulations in a spherical geometry, which puts an upper bound on $Ra_c$ that is similar to what we find in a Cartesian geometry. 
}

\textcolor{RED}{
We expect that the typical size of upwellings will retain a $Ra^{-1/2}$ scaling in spherical geometry, because the balance in (\ref{eq:balance_plume_size}) that gives this scaling remains the same: 
lateral second derivatives must be of order $Ra$ for orthoradial diffusion to be in balance with heat production and radial advection. 
As a consequence, our prediction for the typical buoyancy-flux and hydrothermal-velocity scales should still hold in a spherical geometry.
}

\textcolor{RED}{
Nevertheless, there will inevitably be differences between the flow in a Cartesian and spherical geometry. For example, the ratio of surface area to volume is different, which affects the energy-conservation equation for the time-averaged radial heat flux  $\mbf{\nabla}\cdot (\overline{J} (r) \mbf{e}_r) = 1$ (cf. equation \ref{eq:heat_flux_z}). This constraint imposes $ \overline{J}(r) = r/3$, and the volume-averaged heat flux becomes $\left\langle J \right\rangle  = 1/6$, which is a factor 3 smaller than in a Cartesian geometry. 
As a consequence, we expect heat-flux anomalies to either have lower amplitudes or be sparser. 
This simple analysis suggests that the Cartesian geometry might give an upper bound on the buoyancy flux and the hydrothermal velocity induced in the ocean compared to the real spherical case. 
}

\textcolor{RED}{
\subsubsection*{Depth-dependence of heat production and gravity}
}

The model developed here also neglects any vertical variations of gravity and \textcolor{RED2}{volumetric} heat production, and we briefly explore their possible importance here with the aid of a few additional simulations.
\textcolor{RED2}{
Volumetric heat production on Enceladus decreases away from the center to become negligible close to the surface \cite{choblet_powering_2017}, although the decreases remains sufficiently slow for the heat-production averaged over the spherical shell to increase with radius.
In the uniform heating case, we have found that the structure of the flow is governed by a local balance between advection, diffusion and volumetric heat production. 
We anticipate such a balance to remain at play when heat production varies with depth which makes the volumetric heating variations more relevant to the dynamics than those of the shell-averaged heating.
}
\textcolor{RED2}{Therefore,} to test the effect of vertical variations of heat production, we have carried out simulations with a \textcolor{RED2}{decreasing} source term $q(z) = 2(1-z)$ in the advection-diffusion equation that retains the same spatial average as in the uniform case. 
A typical snapshot is shown in figure \ref{fig:Ra360_gz_snapshot} along with the mean vertical temperature profile, $\overline{\theta}$.
They both show very little difference from the uniform-heating case (see figure \ref{fig:Ra100_field}), suggesting that vertical variation in heat-production does not play an important dynamical role, at least in Cartesian geometry. 

\textcolor{RED}{
The picture is slightly changed when we consider uniform heating but with a depth-dependent gravitational field. In a uniform-density planetary core, we expect $g$ to increase linearly with radius, and so we carried out a few example simulations in which the dimensionless gravity is $g=z$ (i.e. gravity is normalised by its surface value). The effect of this on the equations is to add a factor of $z$ in front of the temperature in the dimensionless version of Darcy's law (see equation (\ref{eq:dimensionless_equations})). 
A snapshot of the temperature field (figure \ref{fig:Ra360_gz_snapshot}) reveals that plumes are narrower and less numerous than in the homogeneous case. 
As a consequence, the maximum advective flux carried by the plumes is roughly double that of the homogeneous case. 
The average temperature profile is also strikingly different: weaker gravity at depth makes advection inefficient as a means of evacuating heat, resulting in larger temperature at the bottom of the domain. 
However, because plumes are thinner, lateral diffusion is enhanced and the plume temperature decreases as they rise. 
As a consequence, the hot-spot temperature at the upper surface remains similar to the modulated case, \ie $\mathrm{max}(\theta(z=1)) \sim 4$. 
}

\textcolor{RED}{
Note that the variations with $z$ of gravity and heat production do not affect the scaling $Ra^{-1/2}$ governing the size of the plumes (and hence the hydrothermal velocity). 
Since the dimensionless gravity  and volumetric heating remain at most order 1, the balance between horizontal  diffusion, heat production and vertical advection still holds in the same way as in the uniform case.
In fact, the asymptotic expansion of section \ref{sec:asymptotic_plume} can be reworked with $z$-dependent gravity and volumetric heating down to equation (31) without affecting the hierarchy between each term. 
Therefore, the scaling laws we have derived in the preceding sections are robust to these additional physical ingredients. 
}

\textcolor{RED}{
We also briefly considered the effect of depth-dependence in the modulation amplitude $\Delta q$ for the results presented in section \ref{sec:heating_large_scale_modulation}, since the modulation should increase with depth, being almost negligible near the core \cite{choblet_powering_2017}. We carried out simulations with $\Delta q = z$, which retains the same average modulation as previously. Results showed little change from those discussed in section \ref{sec:heating_large_scale_modulation}: a mean flow drags small-scale plumes towards the areas with larger volumetric heating which causes the flow to be unsteady.
The strong pulsatility with bursts of heat flux observed at $Ra=360$ in figure \ref{fig:flux_time_mod} are weaker, however, with the time series of $\overline{w\theta}(z=1)$ being more similar to the homogeneous case. 
}

\subsection{Quantification of convection in Enceladus' core}

\textcolor{RED}{
The preceding discussion suggests that the simplified two-dimensional Cartesian model of internally heated porous convection produces scaling laws, at least in terms of orders of magnitude, provide a reasonable description of the flow and the hydrothermal activity inside icy moons. Here we apply our results to Enceladus. 
}
%

\subsubsection*{Physical properties of Enceladus' core}

To characterise convection inside Enceladus' core, and to compare our results to existing literature, we use the same physical parameters as in \cite{choblet_powering_2017}.
A set of fixed physical constants that are relevant to characterise heat transport are given in table \ref{tab:physical_constants}. 
We reproduce the process used in \cite{choblet_powering_2017} and do not precisely specify the permeability $k$ and internal heat production $Q_V$ values \textcolor{RED}{on which the uncertainty is the largest.}
Instead, we consider that $k$ may range from $10^{-15}$ m$^2$ to $10^{-12}$ m$^2$ and that the tidal heating is between $10$ GW and $40$ GW. 
(The lower bound is directly inferred from the heat flux measurement at the South Pole of Enceladus \textcolor{RED}{\cite{spencer_cassini_2006,spencer_plume_2018}.})
Therefore, we draw maps of the behaviour of the system keeping the parameters of table \ref{tab:physical_constants} constant and varying both $k$ and $Q_V$. 

\begin{table}
\centering
\begin{tabular}{|l|l|}
\hline
Core radius ($h$) & 186 km \\
Water density ($\rho_0$) & $1.0 \times 10^{3}$ kg.m$^{-3}$ \\
Matrix density ($\rho_m$) & $2.8 \times 10^{3}$  kg.m$^{-3}$\\
Water heat capacity ($c_0$) & $4.1 \times 10^{3}$ J.K$^{-1}$.kg$^{-1}$\\ 
Matrix heat capacity ($c_m$) & $1.0 \times 10^{3}$ J.K$^{-1}$.kg$^{-1}$\\ 
Water conductivity ($\lambda_0$) & $0.6 \times 10^{3}$ W.K$^{-1}$.m$^{-1}$\\
Matrix conductivity ($\lambda_m$) & $2.8 \times 10^{3}$ W.K$^{-1}$.m$^{-1}$\\
Water thermal expansion ($\alpha$) & $1.2 \times 10^{-3}$ K$^{-1}$ \\
Kinematic viscosity ($\nu$) & $1 \times 10^{-6}$ m$^2$.s$^{-2}$ \\
Thermal diffusivity ($\kappa$) & $6 \times 10^{-7}$ m$^2$.s$^{-2}$\\
Porosity ($\varphi$)& 0.20 \\
Gravity ($g$) & 0.1 m.s$^{-2}$ \\\hline
\end{tabular} 
\caption{A summary of the bulk physical parameters used to transpose our idealised study to the case of Enceladus' core, adapted from \cite{choblet_powering_2017} (see in particular the Supplementary Material of that paper).
Note that the modified porosity is $\overline{\varphi}=0.76$. 
}
\label{tab:physical_constants}
\end{table}

\subsubsection*{The Rayleigh number inside Enceladus}

\begin{figure}
\centering
\includegraphics[width=\linewidth]{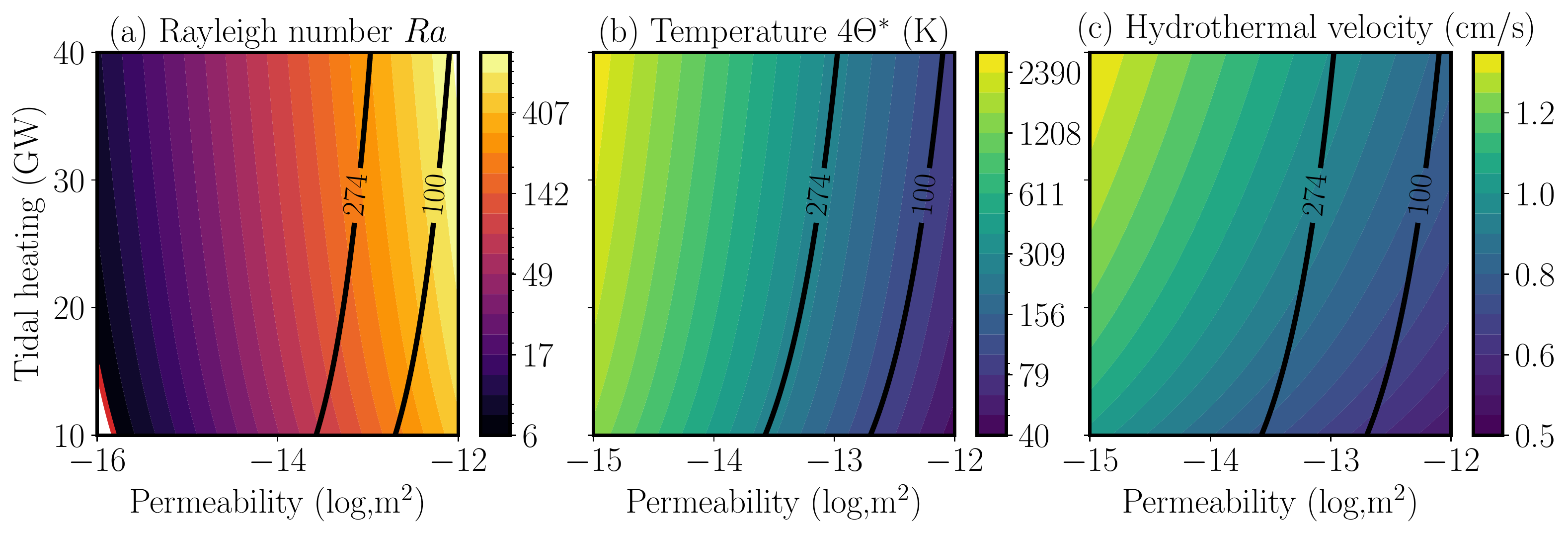}
\caption{\textbf{(a)} the Rayleigh number as a function of the permeability and the tidal heating. 
The red line marks the onset of convection for the homogeneous heating case.
\textbf{(b)} Typical maximum dimensional temperature $4 \Theta^*$ (in Kelvin) (see (\ref{eq:temperature_scale_rayleigh})) inside the porous core of Enceladus. 
\textcolor{RED}{
\textbf{(c)} Typical hydrothermal velocity obtained from the buoyancy flux at the bottom of the ocean as a function of permeability and tidal heating. }
On each panel, the 274 K isotherm gives the liquid-vapour transition \textcolor{RED}{at the hydrostatic pressure of core--ocean boundary, which represents a rough upper bound on the temperature for the model to remain valid with respect to phase change.}
The 100 K isotherm gives \textcolor{RED}{an estimate of the maximum temperature} derived from geochemical measurements \cite{sekine_high-temperature_2015,hsu_ongoing_2015}. }
\label{fig:Rayleigh_number}
\label{fig:temperature_scale_rayleigh}
\end{figure}

As explained in section \ref{sec:scaling_the_problem}, the overall behaviour of the system depends only on one dimensionless parameter, the Rayleigh number, defined in (\ref{eq:Rayleigh_number}), which which is a power law of both $k$ and $Q_V$.
The map of the possible values of the Rayleigh number inside the core of Enceladus is given in figure \ref{fig:Rayleigh_number}. 
In the range of values of $k$ considered in \cite{choblet_powering_2017}, the system is always unstable to convection, although $Ra$ does not reach very high values \textcolor{RED}{ and remains below $1000$}. 

\textcolor{RED}{
\subsubsection*{Maximum temperature}
}
%
%
We have shown in section \ref{sec:advective_heat_transport} that heat transport is mostly advective, even at values of the Rayleigh number that are close to the onset of the instability.
\textcolor{RED}{
In such a regime, the dimensionless temperature takes $O(1)$ values, with a maximum of about $4$ in the case of horizontally modulated heat production (see for instance the snapshots of figure \ref{fig:heterogeneous_heating_snapshots}) in reached at the core of the plumes. 
Hence, $4 \Theta^*$ is a good proxy for the maximum temperature difference between the ocean and the core of Enceladus, with $\Theta^*$ the temperature scale defined in section \ref{sec:scaling},
}
%
%
%
\begin{equation}
\label{eq:temperature_scale_rayleigh}
 \Theta^* ~=~\frac{\kappa \nu}{k \alpha g h} Ra~.
\end{equation}
\textcolor{RED}{The maximum temperature difference $4 \Theta^*$} is shown in figure \ref{fig:temperature_scale_rayleigh} and,  
depending on the parameters, it ranges from 40 K to 3000 K.
\textcolor{RED}{
The computation of the maximum temperature difference allows us to determine the limit of validity of our model which does not include phase change of water from liquid into vapor.
On the one hand, according to the snapshots of figures \ref{fig:Ra100_field} and \ref{fig:heterogeneous_heating_snapshots}, the core temperature of plumes is almost constant with height.
On the other hand, the boiling point of water is an increasing function of pressure and depth. 
As a consequence, the maximum temperature temperature allowed in our model is given by the boiling point of water at the core--ocean boundary. 
Assuming that the core lies below $60$ km of liquid and solid water with density $ 1 \times 10^{3}$ kg.m$^{-3}$ and constant gravity, a crude estimate for the pressure is $6$ Mpa at the core--ocean boundary.
The corresponding boiling point for pure water is $547$ K \cite{haynes_crc_2012}.
If we assume the ocean to be well-mixed and made of pure water, its averaged temperature should be similar to the fusion temperature of ice, that is, $273$ K \cite{haynes_crc_2012}.
Hence, we show in figure \ref{fig:Rayleigh_number} the isotherm 274 K which gives a crude upper bound on the validity of the single-phase model that is used here and in the literature dealing with internal models of icy moons, 
although other additional ingredients (composition, variable gravity) may shift this upper bound in a way that remains to be determined.
}

Note that \citeA{hsu_ongoing_2015} have shown via the ice plume composition that the water flowing inside Enceladus has been in contact with rocks at a temperature of about $90^\circ$ C. 
\textcolor{RED}{
We show in figure \ref{fig:Rayleigh_number} the location of where the maximum temperature difference reaches $100$ K which roughly corresponds to this important constraint on the maximum temperature.
}
Our idealised model suggests a constraint on the permeability of Enceladus' core of \textcolor{RED}{$10^{-13}$ to $10^{-12} $ m$^2$}, for the range of tidal heating considered.  

\textcolor{RED}{
\subsubsection*{Hydrothermal velocity in Enceladus' ocean}
}

\textcolor{RED}
{
Using the law (\ref{eq:buoyancy_2d}) governing the buoyancy flux $B$ driven by porous convection in the ocean, we compute the typical hydrothermal velocity $B^{1/3}$ in Enceladus' ocean.
}
With dimensionless heat flux $w \theta \sim 10$ in the core of the thermal anomalies, the typical hydrothermal velocity is found to be about 1 cm/s, no matter what the permeability or the tidal heating are (see figure \ref{fig:Rayleigh_number}c). 
This value is in agreement with the typical velocity found by \cite{choblet_powering_2017} with different scaling arguments relying on the power anomaly advected to the ocean floor. 
\textcolor{RED}
{
As a consequence, for a subsurface ocean whose thickness is of the order $10-30$ km \cite{thomas_enceladuss_2016-1}, the expected turn-over timescale is of the order of a month, at most.
}

\textcolor{RED}{
\subsubsection*{Typical velocity and temporal variability}
}

The typical velocity scale $U^*$ of the flow in the core is given by a diffusive velocity $\kappa/h$ augmented by a factor $Ra$, that is:
\begin{equation}
U^* = \frac{\kappa}{h} Ra ~.
\end{equation}
The diffusive velocity scale amounts to 0.1 mm.yr$^{-1}$, and because $Ra$ does not exceed \textcolor{RED}{$10^3$}, the Darcy flux remains below $10$ cm.yr$^{-1}$.
The hydrothermal activity at the bottom of Enceladus' ocean is therefore very different from the what is commonly observed \textcolor{RED}{at the bottom of the} Earth's oceans, where typical Darcy fluxes are rather of the order of a few meters per year ($10^3$ times larger).
This difference is largely due to the much weaker gravitational acceleration in Enceladus.
Consequently, the convective time scale $\tau$ is:
\begin{equation}
\tau = \overline{\varphi} \frac{h}{U^*} = \overline{\varphi} \frac{h^2}{\kappa}  Ra^{-1} \simeq 0.8\, \mbox{Gy} \times Ra^{-1}~,
\end{equation} 
The typical variability timescale, for instance for the flux at the top boundary (see figure \ref{fig:flux_time_mod}) is thus at least 1 million years. 
It is a very slowly evolving system \textcolor{RED}{compared to the turnover timescale of the subsurface ocean, or to the timescale of human observations.}
\textcolor{RED}{
In our simulations, we have observed bursts in the convective activity that give rise to a 40-50\% increase in the average heat flux at the surface of the core that may last for a few million years.
These bursts correspond to more active plumes that would cause enhanced hydrothermal activity inducing preferential erosion of the ice shell above. 
One may thus speculate that, in the past, intense plumes similar to the one at the south pole of Enceladus could have been active at other locations.
Such a hypothesis might explain the existence of older tectonised terrains at the surface of Enceladus \cite{crowwillard_structural_2015}. 
Lastly, localised bursts could also be at the start of the runaway mechanism proposed by \cite{choblet_powering_2017} to explain the asymmetry between the north and south poles of Enceladus: a thinner ice crust locally enhances tidal heating, which in turns enhances ice erosion.
}

\section{Conclusion and discussion}

Throughout this article, we have explored heat transport in a fluid-saturated, internally heated porous medium with an idealised Cartesian model. 
Our set-up is based on an idealisation of the model of \citeA{choblet_powering_2017} describing the tidally-driven hydrothermal activity in the interior of Enceladus.
The behaviour of the system is governed by a single dimensionless number, the Rayleigh number $Ra$, which is an increasing function of both the permeability and the internal heat production.
With the combination of numerical simulations and mathematical analysis, we have derived general laws governing hydrothermal activity driven by volumetric heating. 
We have shown that heat transport in the porous medium is governed by advection.
In this regime, the temperature difference between the porous matrix and the pure fluid ocean scales like $Ra$. 
This scaling enables use to constrain the plausible range of values for the permeability of Enceladus' core. 
According to \citeA{hsu_ongoing_2015}, the temperature scale should be at most 100 K. 
For values of tidal heating that are consistent with the heat flux measurement at the surface of Enceladus, our scaling indicates that the permeability should be around $10^{-13}-10^{-12}$ m$^2$.
In our simulations, we have reproduced the observation drawn from the simulations of \citeA{choblet_powering_2017} that the upwelling zones tend to narrow as the Rayleigh number is increased, and that they concentrate where internal heating is the most intense. 
Our simulations show that the typical plume size follows a $Ra^{-1/2}$ power law, which is imposed by a balance between vertical advection and horizontal diffusion of heat. 
This law governing the size of heat flux anomalies at the bottom of the ocean of Enceladus compels the typical buoyancy flux injected into the ocean to be proportional to $Ra^{3/2}$.
Over the range of tidal heating and permeability that are consistent with observational data, we have found that the typical hydrothermal velocity in the ocean of Enceladus is about 1 cm/s. 
Despite the idealisation of our model, such an estimate is consistent with the one derived by \citeA{choblet_powering_2017} from an estimate of a typical heat flux anomaly. 
The model used here has also helped us to highlight the underpinning of heat transport in an internally heated porous layers.
In particular, we have shown that the heat-transport efficiency, which has been characterised via a generalised Nusselt number, has the same scaling as the classical Rayleigh-B\'enard convection in porous media \cite{otero_high-rayleigh-number_2004,hewitt_ultimate_2012,hewitt_high_2014}. 
\textcolor{RED}{
Despite the highly idealised nature of our approach, we have argued that the scaling laws found for the typical size of thermal anomalies, the time-variability and the hydrothermal activity are also expected in spherical geometry, and are robust to the inclusion of additional ingredients such as vertical variations of heat production and gravity. 
%
%
These scaling laws could thus be applied to the other small icy moons of the Solar System, in particular those of Saturn's E ring, whose internal structure is similar to Enceladus' \cite{nimmo_ocean_2016}.
Although Enceladus is the only one showing signs of present internal activity, these other bodies could have been active in the past.
Our study thus paves the way for more systematic understanding of the thermal evolution of these bodies.
It could also apply to larger icy moons such as Europa where the ocean is in contact with a rocky mantle that is internally heated by radiogenic decay.
}

%
%
%
%
%
\textcolor{RED}{
Of course, the simple scaling arguments contained here are not a substitute for a detailed investigation of the idealised problem in a spherical geometry, which would be a useful future extension to this work. Such a study would give a clearer picture of the flow structures in a spherical geometry, as well providing more quantitative predictions of hot-spot widths, time-variability and strength. 
Beyond these geometrical considerations, there are also other effects that have not been discussed here or in the existing literature that could lead to significant changes in convective heat transport.
}
One major simplification of all models of porous planetary interiors is the assumption of homogeneous and isotropic permeability $k$. 
Since the core of small icy satellites such as Enceladus' is an aggregate of heterogeneous material, their permeability is unlikely to be uniform.
It is not even clear whether coarse-grained modelling based on the assumption of strong confinement, that is, Darcy's law, is entirely relevant for the core of icy satellites, although, as we've seen, Darcy's law with small permeability is consistent with observed data.
However, we do not believe any significant progress can be achieved in these directions without further constraining the core's small-scale structure.
In addition, the model we consider completely discards flows that are directly driven by the periodic tidal distortion. 
Although the tidal deformation field  is purely incompressible in continuous media, mean flows analogous to Stokes drift may result from the periodic motion of the porous matrix. 
Whether deformation-driven flows are comparable to buoyancy-driven flows remains to be quantified.  
Lastly, there is a need to clarify the behaviour of the system at the top of the porous core and the coupling between the porous layer and the above ocean. 
We have stated in the second section that the two possible thermal boundary conditions used here (imposed temperature or free temperature in the upwellings) are the two end-members of the behaviour of the fluid at the interface.
The imposed temperature condition could represent a very slow porous layer lying underneath a very well mixed ocean.
\textcolor{RED}{
This situation could be relevant to the case of Enceladus and other icy moons as the Darcy flux ($\sim$ 1 cm.yr$^{-1}$) is very small compared to the hydrothermal velocity ($ \sim 1$ cm.s$^{-1}$).}
In this configuration, the water coming out of the core is at the same temperature as the ocean and is neutrally buoyant; there is then no hydrothermal activity in the sense of what we know at the bottom of the Earth's ocean.
Nevertheless, it is associated to a diffusive heat flux anomaly on the subsurface ocean's floor which is likely to drive convection and mixing in the ocean.
The observed chemical signature of contact with silicate rocks at high temperature \cite{hsu_ongoing_2015} could very well happen below the thin thermal boundary layer at the top of the core. 
Moreover, current thermal evolution models of icy moons rely on parametrisation of hydrothermally-driven convection in the sub-surface ocean that are based on the classical Rayleigh-B\'enard problem \cite{travis_whole-moon_2012,travis_keeping_2015}.
It is, however, not clear at all whether such parametrisation actually applies to the present system where ocean convection is driven by strong and localised heterogeneities of either the advective of the diffusive heat flux at the bottom boundary
In short, it remains difficult to produce a definitive statement about the thermal structure of the subsurface ocean without a careful study of the coupled system with two very different typical evolution timescales for each medium.

\acknowledgments
The authors acknowledge support from the 2018 WHOI  GFD  program,  supported  by  the US  National  Science Foundation  (award  no.  1332750)  and  the  Office  of  Naval  Research, were most of this research was carried out.
TLR is supported by the Royal Society through a Newton International Fellowship (Grant reference NIF\textbackslash R1\textbackslash 192181).
The source codes and output data are available on Figshare \cite{le_reun_internally_2020-1}.


%


\end{document}